\documentclass[10pt,journal]{IEEEtran}
\usepackage{cite}
\usepackage{amsmath,amssymb,amsfonts}
\usepackage{algorithmic}
\usepackage{url}
\usepackage{graphicx}
\usepackage{textcomp}
\usepackage{xcolor}
\usepackage{amsmath,lipsum}
\usepackage[ruled]{algorithm2e}
\usepackage{graphicx}
\usepackage{float}
\usepackage{subfigure}
\usepackage{multirow}
\usepackage{multicol}
\usepackage{arydshln}
\usepackage{cite}
\usepackage{stfloats}
\usepackage{booktabs}
\usepackage{caption}
\usepackage{makecell}

\newtheorem{theorem}{Theorem}

\begin{document}

\title{Dynamic Semantic Compression for CNN Inference in Multi-access Edge Computing: A Graph Reinforcement Learning-based Autoencoder}

    \author{Nan Li,~\IEEEmembership{Student Member,~IEEE,} Alexandros Iosifidis,~\IEEEmembership{Senior Member,~IEEE and} Qi Zhang,~\IEEEmembership{Senior Member,~IEEE}
\thanks{This work is supported by Agile-IoT project (Grant No. 9131-00119B) granted by the Danish Council for Independent Research. Part of this paper is accepted by IEEE ICC 2023 \cite{Nan2023ICC}.}
\thanks{N. Li, A. Iosifidis, and Q. Zhang are with the Department
of Electrical and Computer Engineering, Aarhus University,
Finlandsgade 22, 8200, Denmark, DIGIT and
(email: lnzyy170320@gmail.com; ai@ece.au.dk; qz@ece.au.dk).}}

\markboth{}%
{Shell \MakeLowercase{\textit{et al.}}: A Sample Article Using IEEEtran.cls for IEEE Journals}

\IEEEpubid{}

\maketitle

\begin{abstract}
This paper studies the computational offloading of CNN inference in dynamic multi-access edge computing (MEC) networks. To address the uncertainties in communication time and computation resource availability, we propose a novel semantic compression method, autoencoder-based CNN architecture (AECNN), for effective semantic extraction and compression in partial offloading. 
In the semantic encoder, we introduce a feature compression module based on the channel attention mechanism in CNNs, to compress intermediate data by selecting the most informative features. In the semantic decoder, we design a lightweight decoder to reconstruct the intermediate data through learning from the received compressed data to improve accuracy.
To effectively trade-off communication, computation, and inference accuracy, we 
design a reward function and formulate the offloading problem of CNN inference as a maximization problem with the goal of maximizing the average inference accuracy and throughput over the long term. To address this maximization problem, we propose a graph reinforcement learning-based AECNN (GRL-AECNN) method, which outperforms existing works DROO-AECNN, GRL-BottleNet++ and GRL-DeepJSCC under different dynamic scenarios. This highlights the advantages of GRL-AECNN in offloading decision-making in dynamic MEC. 
\end{abstract}

\section{Introduction}
\IEEEPARstart{T}{he} widespread adoption of Internet of Things (IoT) devices, has paved the way for developing real-time and context-aware applications, such as autonomous driving and augmented reality. These devices generate enormous volumes of data, necessitating efficient processing and inference capabilities. However, the limited computational resources and constrained bandwidth on IoT devices pose significant challenges in performing \textit{local computing}, especially for computationally intensive convolutional neural networks (CNNs) that require massive multiply-accumulate operations \cite{Guo2016}. To perform the computation-demand and memory-required CNN inference task within a stringent deadline, a common approach is to compress and prune CNN topology thereby reducing the computational operations. However, over-pruning CNNs may cause severe accuracy degradation. 

To mitigate this issue, edge computing has emerged as an efficient approach, enabling IoT devices to fully offload computational tasks (i.e., \textit{full offloading}) to edge servers (ESs) through wireless channels \cite{008}. However, the fluctuations in communication time caused by stochastic wireless channel states may introduce inherent uncertainty in the communication time, resulting in varying and unpredictable communication delays \cite{Tran2019TVT}. In addition, the varying size of inference tasks generated by IoT devices adds further variability, contributing to the overall uncertainty in communication delays. Consequently, the uncertainty of communication time directly affects the available time budget for performing the computation, which may lead to task failure when the computation cannot be completed within the deadline. Furthermore, the computational resources of each ES are usually shared by multiple IoT devices, resulting in dynamic changes in resource availability \cite{Chang2021TII}. This unpredictable computation resource exacerbates the uncertainty in computation time, further increasing the likelihood of tasks failing to meet the deadline. 

To strike a balance between communication and computation, dynamic offloading methods have been proposed to optimize the offloading decision-making process \cite{Guo2016, Huang2020}. However, when communication takes too much time or the available computation resources at ESs are insufficient, meeting stringent deadlines by running the entire pre-trained CNN model on ES becomes challenging. As such, dynamic neural networks such as 
skipping layers \cite{ Wang_2018_ECCV}, kernel filters \cite{gao2018dynamic} and early-exits \cite{Surat2016ICPR}, modify the CNN architecture thereby allowing dynamic inference time at the expense of inference accuracy. However, dynamic neural networks still face challenges in meeting strict time constraints due to uncertain communication and computation time, potentially resulting in significant degradation of inference accuracy. These limitations have driven the development of other alternatives, among which \textit{split computing} (i.e., partial offloading) has shown promise in striking a balance between communication and computation \cite{Jeong2018}. However, most existing works on \textit{split computing} primarily focus on model splitting, and less attention was paid to the compression of intermediate feature \cite{BottleNet++}. 

\begin{figure}[t]
    \centering
    \includegraphics[width=0.4 \textwidth]{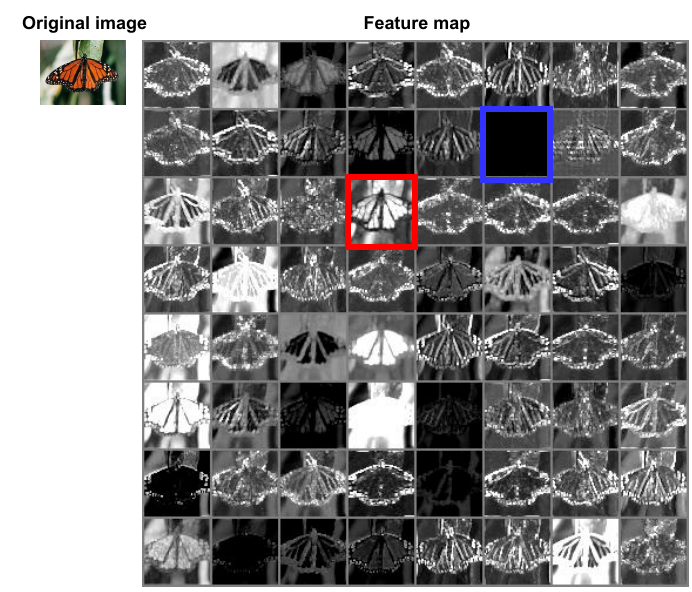}
    \caption{The 1st CL's output feature maps in ResNet-50.
    The blue one is almost useless for inference, while the red one has enough information to be used to generate the rest.}
    \label{fig:feature}
\end{figure}

In general, a well-trained CNN model often contains redundant features that are not essential for performing an inference task \cite{Han_2020_CVPR}, and not all of these features play the same role therein (as shown in Fig. \ref{fig:feature}), i.e., different features have varying degrees of importance in making predictions \cite{Woo_2018_ECCV}. Therefore, under poor wireless channel conditions, it is desirable to prune less important features to reduce communication overhead thereby meeting the deadline. This idea aligns with the emerging paradigm of semantic communication, which aims to extract the ``meaning" of information to be transmitted at a transmitter and successfully interpret the received semantic information at a receiver \cite{Luo2022}. Semantic compression can be used to extract and utilize semantic information to compress the intermediate tensor in the early layers in partial offloading and optimize the communication process. For example, in image classification tasks, not all the features but only the local features (e.g., pixels) of the image directly relevant to the classification are transmitted thereby reducing the communication overhead \cite{Brendan2011SemCom}. Motivated by the fault-tolerant property of CNNs, Shao et al. \cite{BottleNet++} proposed BottleNet++, which used a CNN-based encoder to resize the feature dimension. Similarly, Jankowski et al. \cite{Jankowski2021JSAC} proposed DeepJSCC to compress the intermediate feature by using a CNN-based encoder. However, directly resizing feature dimensions may compromise the effective representation of the semantic information in the features and result in accuracy degradation.

In wireless edge computing systems, time-varying wireless channel states and available computing resources significantly impact the optimal decision-making process for offloading tasks, especially in multi-access edge computing (MEC) networks. In MEC, one of the major challenges is the joint optimization of computing paradigms (i.e., \textit{local computing}, \textit{full offloading} or \textit{split computing}), wireless resource allocation (e.g., transmission power, transmission size of intermediate semantic information) and inference accuracy. This optimization problem involves ternary offloading variables and is typically formulated as a mixed integer programming (MIP) problem \cite{Huang2020}, which can be solved using dynamic programming and heuristic local search methods. 
However, these approaches either suffer from prohibitively high computational complexity or require a considerable number of iterations to converge to an optimal solution, making them impractical for real-time offloading decisions in time-varying wireless channels \cite{Huang2020}. Reinforcement learning (RL) is a holistic learning paradigm that interacts with the dynamic MEC to maximize long-term rewards. Li et al. \cite{Li2018} proposed to use deep RL (DRL)-based optimization methods to address dynamic computational offloading problem. However, applying DRL directly to the problem is inefficient in a practical deployment because it typically requires many iterations to search an effective strategy for unseen scenarios. Huang et al. \cite{Huang2020} proposed DROO to significantly improve the convergence speed through efficient scaling strategies and direct learning of offloading decisions. However, the DNN used in DROO can only handle Euclidean data, which makes it not well suitable for the graph-like structure data of MEC. In addition, all the above methods do not provide dynamic inference, which is lack of flexibility in making good use of any available computation resource under stringent latency. 

In this paper, we propose an adaptive semantic compression technique to address the challenges associated with executing computationally intensive CNN inference tasks in MEC. Our approach leverages the advancements in semantic communication to achieve efficient CNN inference offloading while maintaining inference accuracy. The main contributions are summarized as follows:

\begin{itemize}
\item \textit{Semantic Encoder:} We design a feature compression module based on the \textit{channel attention} (CA) method in CNNs to quantify the importance of channels in the intermediate tensor. By utilizing the statistics of channel importance, we can calculate the importance of each channel, enabling intermediate tensor compression by pruning channels with lower importance. Furthermore, we employ entropy encoding to remove statistical redundancy in the compressed intermediate tensor, further reducing the communication overhead.

\item \textit{Semantic Decoder:} We design a lightweight \textit{feature recovery} (FR) module that employs a CNN to learn and recover the intermediate tensor from the received compressed tensor. This process enhances inference accuracy by effectively reconstructing the compressed tensor.

\item \textit{Reward Function and Optimization:} We define a reward function that strikes a balance between communication, computation, and inference accuracy. The CNN inference offloading problem is formulated as a maximization problem to optimize the average inference accuracy and throughput over the long term under the constraints of latency and transmission power.

\item \textit{Graph Reinforcement Learning (GRL)-based Autoencoder:} To address the challenges posed by stochastic available computing resources at ESs and uncertainties in communication time, we propose GRL-AECNN to ensure that the inference task is completed within the given time constraints by leveraging the capacity of reinforcement learning and graph convolutional network (GCN). 

\item \textit{Performance Evaluation:} We employ a step-by-step approach to fasten the training process \cite{He_2016_CVPR}. Experimental results demonstrate that GRL-AECNN achieves better performance than the existing works DROO-AECNN, GRL-BottleNet++ and GRL-DeepJSCC under different dynamic scenarios, which demonstrates the effectiveness of GRL-AECNN in offloading decision-making. 
\end{itemize}

The remainder of this article is organized as follows.  The
system model is presented in Section \ref{section:system_model}. Section \ref{Section:AECNN} describes the proposed AECNN architecture for CNN inference offloading. In Section \ref{problem}, the CNN inference offloading problem is modeled as a maximization problem. In Section \ref{method}, GRL-AECNN method is proposed to solve the optimization problem  
The simulation results are presented and discussed in Section \ref{Performance}, and the conclusions are drawn in Section \ref{Conclusion}. 
The notations used in this paper are listed in Table \ref{tab:notations}.

\section{System Model}\label{section:system_model}
\begin{figure}[t]
    \centering
    \includegraphics[width=0.4\textwidth]{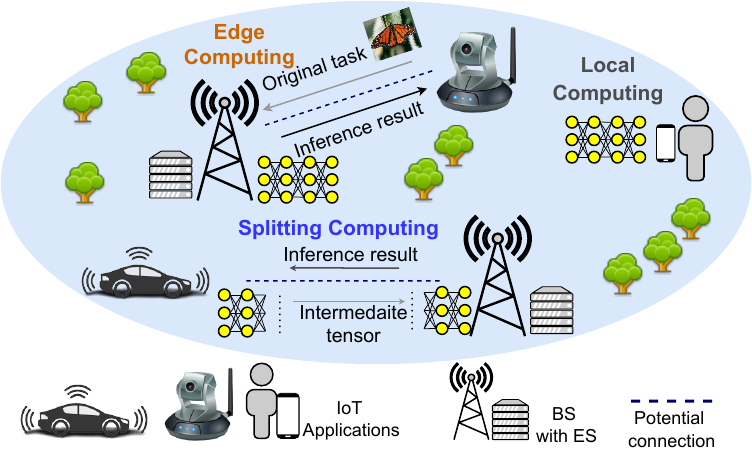}
    \caption{An example of task offloading in an MEC network.}
    \label{fig:mec}
\end{figure}

We consider a dynamic MEC network composed of $U$ IoT devices and $S$ ESs, as illustrated in Fig. \ref{fig:mec}. The set of IoT devices and ESs are denoted as $\mathcal{U}
= \left\{ 1, 2, \cdots, U \right\}$ and $\mathcal{S} = \left\{ 1, 2, \cdots, S \right\}$ respectively. At each timeslot $k \in \mathcal{K} = \left\{1, 2, \cdots, K\right\}$, each IoT device generates a computational task that needs to be processed within a given time constraint. The duration of each timeslot is assumed to be constant and denoted as $\tau$. We mainly focuses on the image classification task and assumes that the computational task utilizes a CNN model $\Omega $ with $L$ convolutional layers (CLs) and several fully-connected (FC) layers. We denote the set of CLs as $\mathcal{L} = \left\{0, 1, \cdots, L \right\}$, where the special layer $0$ represents the initial stage of the CNN computation. To execute the computational task, each IoT device adheres to a ternary computational policy, i.e., \textit{local computing}, \textit{full offloading} or \textit{split computing}. 

\subsection{Task Model}
The parameters associated with the computational tasks at timeslot $k$ are defined as $\mathcal{I}_k \triangleq \{ \left.  \left(d_{u}^k,\sigma_{u}^k\right) \right|\forall u \in \mathcal{U} \} $. 
Here $d_{u}^k$ represents the size of the task generated by IoT device $u$ at timeslot $k$, typically referring to the size of the original image unless otherwise specified in this paper. The parameter $\sigma_{u}^k$ indicates the maximum tolerable latency, ensuring that the latency experienced by each inference task does not exceed $\sigma_{u}^k$. For the sake of clarity, we consider \textit{local computing} and \textit{full offloading} as two distinct cases within the framework of \textit{split computing}, and introduce a binary variable $\alpha_{u,l}^k \in \{0,1\}$ to indicate whether the inference task generated by IoT device $u$ at timeslot $k$ is split at CL $l$. Specifically, $\alpha_{u,0}^k = 1$ means that the entire inference task is fully offloaded to an ES; $\alpha_{u,L}^k = 1$ denotes that the task is performed locally; otherwise, the computation of the inference task is split between IoT device $u$ and an ES, wherein the IoT device offloads the intermediate feature map to the ES after computing the first part locally (i.e., from CL 0 to $l$), and then the ES performs the remaining computation (i.e., from CL $l+1$ to $L$) and send back the result to IoT device. Since each IoT device can only utilize one computation mode to perform the inference task at each timeslot, a feasible offloading policy must satisfy with the following constraint:
\begin{equation}
    \begin{array}{*{20}{c}}
    \sum\limits_{l \in \mathcal{L}}  \alpha_{u,l}^k = 1, &\forall u \in \mathcal{U}. 
        \end{array}
\end{equation}
Additionally, we define a binary variable $\beta_{u,s}^k \in \{0,1\}$ to indicate whether the computation of IoT device $u$ at timeslot $k$ is offloaded to ES $s$, where $\beta_{u,s}^k = 1$ means that the computation of IoT device $u$ is offloaded to ES $s$, and vice versa. Therefore, we have
\begin{equation}
    \sum_{s \in \mathcal{S}}  \beta_{u,s}^k = \left\{ {\begin{array}{*{20}{c}}
        0,&\alpha_{u,0}^k = 1,\\
        1,& {\rm otherwise}.
        \end{array}} \right.
\end{equation}

\renewcommand\arraystretch{1.3}
\begin{table}
    \centering
    \caption{Summary of Notations}
	\label{notations}
 \resizebox{\linewidth}{!}{
    \begin{tabular}{cl}
    \toprule
    \textbf{Notation} & \textbf{Description} \\
    \toprule
    $ \mathcal{L}, \mathcal{U}, \mathcal{S}, \mathcal{K}, \mathcal{M}$   & 
    \makecell[l]{The set of CLs, IoT devices, ESs,  timeslots,  compression ratios }
    \\ 
    $\alpha_{u,l}^k, \beta_{u,s}^k, \gamma_{u,s}^k$    & \makecell[l]{
    The CNN inference offloading decisions at timeslot $k$ }     \\ 
    $\Omega$    & CNN model with $L$ CLs and several FLs \\ 
    $\mathcal{X}_l$   & \makecell[l]{The output feature map of CL $l$ }   \\ 
    $\textrm{FA}_l, \textrm{FM}_l$  & The aggregated features after avg-pooling and max-pooling  \\ 
    $\textrm{WA}_l$   & The attention weight map of CL $l$  \\

    \specialrule{0em}{2pt}{2pt}

    $R_{u,s}^k, B_{u,s}^k$    & The data rate, bandwidth between IoT device $u$ and ES $s$    \\ 
    
    $p_{u,s}^k, g_{u,s}^k $    & The transmission power, channel gain of link between $u$ and $s$ \\

    $\mathcal{I}_k$    & \makecell[l]{The parameters of task with size $d_{u}^k$ and deadline $\sigma_{u}^k$}  \\ 
    
    $D_{u,s}^k$    & \makecell[l]{The transmitted data size of IoT device $u$ }     \\  
    $n_0$    & The background noise power spectral density \\
    $P_u$    & The maximal transmission power of IoT device $u$  \\
    $t_{u,s,k}^{\textit{com}}$    & The communication time of $u$ offloading data to $s$ \\
    $t_{u,i,j}^{cmp}, t_{s,i,j}^{cmp}$   & The computation time from CL $i$ to CL $j$ on $u$ and $s$\\
    $t_{u,k}^{\textit{cmp}}$    & The computing time of $u$ at timeslot $k$  \\ 
    $t_{u,s, k}^{\textit{cmp}}$    & The computing time of $s$ performing $u$'s task   \\ 
    $T_{u,s,k}^{\textit{arr}}$    & The arrival time instant of $u$'s task at $s$   \\ 
    $t_{u,s, k}^{\textit{que}}$    & The queue delay of $u$'s task at $s$   \\ 
    $t_{u,s, k}$   & The total completion time of $u$'s task at $s$   \\ 
    $E_{u,s,k}^{\textit{com}}$    & The energy consumption of $u$ transmitting data to $s$ \\
    $E_{u,l,k}^{\textit{cmp}}$    & The energy consumption of $u$ computing CL $0$ to CL $l$ \\
    $E_{u,s, k}$   & The energy consumption of $u$
    when offloading task to $s$  \\
    \bottomrule
\end{tabular}
}
\label{tab:notations}
\end{table}

When the computation is split at CL $l \in \mathcal{L}\backslash \left\{0, L\right\}$, the intermediate data tends to be a high-dimensional tensor, leading to a potential increase in communication overhead. To address this challenge, we propose a novel semantic compression approach named \textbf{AECNN}, as detailed in Section \ref{Section:AECNN}. \textbf{AECNN} allows to compress the intermediate feature map at a predefined compression ratio $ m \in \mathcal{M}=\left\{1, 2, \cdots, M \right\}$ while maintaining an acceptable level of inference accuracy. To facilitate this compression process, we utilize a binary variable $\gamma_{u,m}^k \in \left\{0,1 \right\}$ to indicate whether the intermediate tensor of IoT device $u$ at timeslot $k$ is compressed using a compression ratio $m $. Therefore, we have 
\begin{equation}
    \sum_{m \in \mathcal{M}}  \gamma_{u,m}^k = \left\{ {\begin{array}{*{20}{c}}
        1,&\alpha_{u,l \in \mathcal{L}\backslash \left\{0,L\right\}}^k = 1 ,\\
        0,& {\rm otherwise}.
        \end{array}} \right.
\end{equation}
Note that compressing the intermediate data may result in a degradation of inference accuracy. Therefore, we use $ \eta_{u,m}^k$ to denote the achieved inference accuracy of the task generated by $u$ at timeslot $k$ when employing the compression ratio $m$.

In general, to perform the computational task within a given deadline, the following three decisions need to be made: to which CL the CNN model should be split, i.e., $\alpha_{u,l}^k$; to which ES an IoT device should offload its tasks, i.e., $ \beta_{u,s}^k $; which compression ratio an IoT device should select, i.e., $ \gamma_{u,m}^k$.

\subsection{Communication Model}
In case that IoT device offloads its entire task or intermediate tensor to an ES, the incurred transmission delay involves delivering of entire inference task or intermediate feature map and its inference result between IoT device and ES. Since the output of the CNN is typically a small-sized value representing the classification or detection result, we do not consider the transmission delay of the feedback in this paper. Consequently, the amount of data transmitted from IoT device $u$ to ES $s$ can be described as follows:
\begin{equation}
    D_{u,s}^k = 
    \left\{ {\begin{array}{*{20}{c}}
        \beta_{u,s}^k d_{u}^k,&a_{u,0}^k = 1,\\
        \frac{4 \beta_{u,s}^k  C_l H_l W_l}{\sum \limits_{m \in \mathcal{M}}  \gamma_{u,m}^k  m},& a_{u,l \in \mathcal{L}\backslash \left\{0,L\right\}}^k = 1,\\
        0,&a_{u,L}^k = 1,
        \end{array}} \right.
\end{equation}
where $C_l$, $H_l$ and $W_l$ are the channel, height and width dimensions of CL $l$'s output feature map $\mathcal{X}_l \in \mathbb{R}^{C_l \times H_l \times W_l}$, respectively. Note that the output tensor $\mathcal{X}_l$ is usually in float32 data type, and the size mentioned above is measured in bytes. 

We assume the uplink channel gain between IoT device $u$ and ES $s$ at timeslot $k$ is denoted as $g_{u,s}^k 
\in \mathcal{G}_k = \{ \left. g_{u,s}^k  \right|\forall u \in \mathcal{U}, \forall s \in \mathcal{S} \} $, capturing the effects of path loss and shadowing fading. Consequently, the uplink transmission data rate from IoT device $u$ to ES $s$ can be expressed as:
\begin{equation}
    R_{u,s}^k = B_{u,s}^k \log_2 \left(1+ \frac{ p_{u,s}^k g_{u,s}^k}{n_0B_{u,s}^k}
     \right), 
\end{equation} 
where $B_{u,s}^k$ is the channel bandwidth allocated to the link between IoT device $u$ and ES $s$, and $n_0$ denotes the noise power spectral density. The transmission power of IoT device $u$ when offloading the entire task or intermediate feature map to ES $s$, $p_{u,s}^k$, should not be greater than its maximal transmission power $ P_{u} $, i.e., $ p_{u,s}^k \leq P_{u} $.

During data transmission, we do not consider the data overhead introduced by the network protocol stack and forward error correction. Therefore, the transmission delay of IoT device $u$ when offloading
its entire task or intermediate feature map to ES $s$ can be expressed as,
\begin{equation}
    t_{u,s,k}^{\textit{com}} =  D_{u,s}^k /R_{u,s}^k. 
\end{equation}

In terms of the energy consumption during data transmission, we do not consider the efficiency of the power amplifier in the antenna and power consumption in the baseband circuit. Therefore, the energy consumption incurred by IoT device $u$ when offloading data to ES $s$ can be represented as
\begin{equation}
    E_{u,s,k}^{\textit{com}} =   t_{u,s,k}^{\textit{com}} p_{u,s}^k. 
\end{equation}

\begin{figure*}[t]
    \centering
    \includegraphics[width= \textwidth]{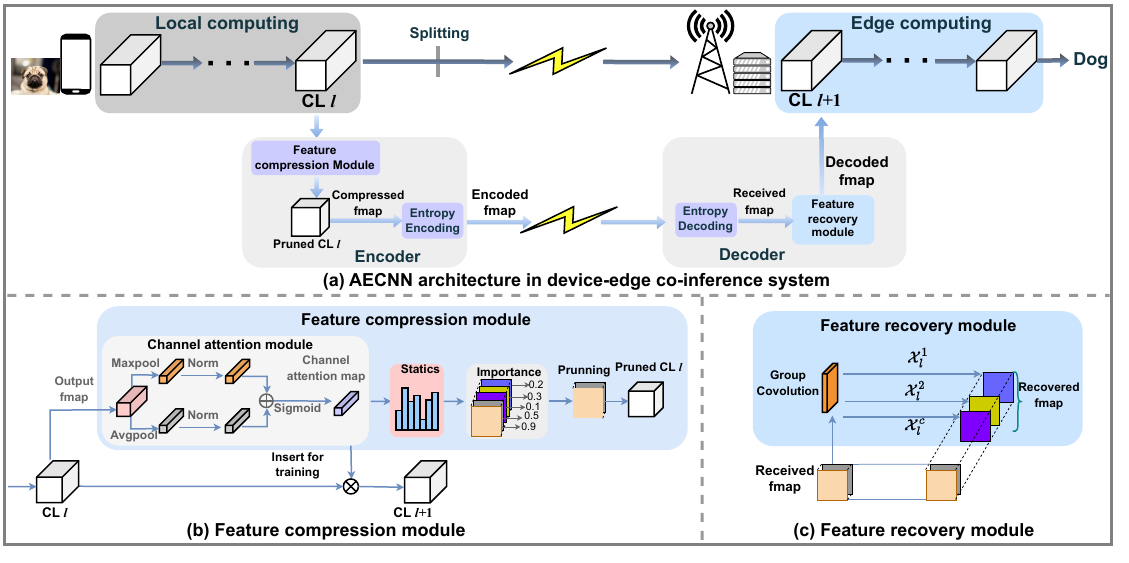}
    \caption{The proposed AECNN architecture in device-edge co-inference system. (a) depicts the overall framework of
    AECNN; (b) shows the design of FC module; and (c) displays the designed FR module by using a CNN with group covolutional layers.}
    \label{fig:AECNN}
\end{figure*}
\subsection{Computation Model}
In CNN, the computation time is specific to the hardware architecture, and can vary based on various factors, including the device, power management techniques, memory access patterns and etc \cite{arxiv:energy_flops}. Therefore, we employ statistical methods to measure the computation time of each layer, and denote the measured computation time from CL $i$ to CL $j$ on IoT device $u$ and ES $s$ as $t_{u,i,j}^{cmp}$ and $t_{s,i,j}^{cmp}$, respectively. 

In \textbf{AECNN}, the feature compression module is needed only during the training phase for the pruned CL $l$ but not in the inference process. Additionally, the computation time required for the lightweight FR module at the ES is so small that can be considered negligible in practice. Therefore, the computation time includes the CNN computation time $t_{u,0,l}^{cmp}$ and the feature encoding time $t_{u,l,m}^{\textit{enc}}$ on the IoT device, as well as the feature decoding time $t_{s,l,m}^{\textit{dec}}$ and CNN computation time on the ES $t_{s,l+1,L}^{cmp}$. Correspondingly, we present the computation time of an inference task on IoT device $u$ and ES $s$ as 
\begin{equation}
    t_{u,k}^{\textit{cmp}} = 
    \left\{ {\begin{array}{*{20}{c}}
        0, &\alpha_{u,0}^k = 1,\\
        t_{u,0,l}^{cmp} + t_{u,l, m}^{enc},& \alpha_{u,l \in \mathcal{L}\backslash \left\{0,L\right\}}^k = 1,\\
        t_{u,0,L}^{cmp},&\alpha_{u,L}^k = 1,
        \end{array}} \right.
\end{equation}
and 
\begin{equation}
    t_{u,s,k}^{\textit{cmp}} = 
    \left\{ {\begin{array}{*{20}{c}}
        \beta_{u,s}^k t_{s,0,L}^{cmp}, &\alpha_{u,0}^k = 1, \\
        \beta_{u,s}^k \left(t_{s,l,m}^{dec} + t_{s,l+1,L}^{cmp} \right), &\alpha_{u,l \in \mathcal{L}\backslash \left\{0,L\right\}}^k = 1,\\
        0,&a_{u,L}^k = 1.
        \end{array}} \right.
\end{equation}
 
Given the energy constraints of IoT devices, we primarily focus on investigating the energy consumption of CNN inference tasks on these energy-constrained IoT devices.
According to \cite{flos_tools}, we denote the number of floating-point operations (FLOPs) that can be performed per Watt per second as $\rho$ and calculate the energy consumption on IoT device $u$ as
\begin{equation}
    \begin{array}{*{20}{c}}        E_{u,l,k}^{\textit{cmp}} = \sum \limits_{i=0}^l \xi_l / \rho   , & \alpha_{u,l}^k =1,
    \end{array}
\end{equation}
where $\xi_i$ represents the FLOPs count of CL $i$. The detailed calculation of the FLOPs count can be found in Appendix \ref{Flops}.

\section{Architecture of AECNN} \label{Section:AECNN}
In this section, we first present an overview of our proposed AECNN architecture. Next, we describe the structural components of our designed feature compression module in the encoder and how to compress the intermediate tensor. Finally, we introduce the designed FR module in the decoder.
\subsection{Overview of AECNN Architecture}
Fig. \ref{fig:AECNN} depicts the overall framework of our proposed AECNN, which consists of an encoder and a decoder. In the encoder, a CA module is designed to assess the statistical importance of channels during inference. This enables the pruning of the channels with low importance based on a predefined compression ratio $\mathcal{M}$. Subsequently, an entropy coding module can be applied to remove the statistical redundancy in the remaining intermediate features. Finally, the decoder first uses the entropy decoding module to decode the received data, and then uses the designed FR module to recover the pruned features from the decoded features. 

\subsection{Feature compression module}
The attention mechanism can effectively improve the classification performance of CNNs by enhancing the representation of features with more important information and suppressing unnecessary information interference \cite{Woo_2018_ECCV}. Channel attention used in CNN usually focuses on evaluating the importance of tensor's channels by paying attention to different channels of the tensor. For example, for CL $l$, each element of the channel attention map $W_l \in \mathbb{R}^{C_l \times 1 \times 1}$ corresponds to a channel's weight of the output tensor $\mathcal{X}_l$. As such, the channels with lower importance can be identified and removed, thereby reducing the size of the intermediate tensor and reducing the communication and computation time on the IoT device. Note that as IoT device knows which channels will be pruned for any pre-defined compression ratio, it only needs to compute the retained channels of the output tensor.

In the previous designs of channel attention \cite{Woo_2018_ECCV}, two FC layers are used to handle the attention weight of the channels. However, this may introduce two drawbacks. First, the reduction of channel dimensionality for saving computation overhead may have side effects on the prediction of channel attention. Second, the learned channel attention by FC is intrinsically implicit, resulting in unknowable behavior of neuronal output. To address these issues, normalization can yield competition or cooperation relationships among channels, using fewer computation resources while providing more robust training performance \cite{Yang_2020_CVPR}. Motivated by the above, we design a CA module, i.e., a global max-pooling layer and a global average-pooling layer with normalization, and insert it into the original CNN model after the splitting point $l$, and then train the resulting network to generate the importance value of each channel, as shown in Fig. \ref{fig:AECNN}(b). 

Since the calculation of the global avgpooling layer and global maxpooling layer are similar, we hereby take the avgpooling layer as an example. The aggregated features after the avgpooling layer can be represented as
\setcounter{equation}{10}
\begin{equation}
    \textrm{FA}_l = \textit{AvgPool} \left(\mathcal{X}_l\right),
\end{equation}
where $\textrm{FA}_l \in \mathbb{R}^{C_l \times 1 \times 1}$. And then the aggregated features $\textrm{FA}_l$ is normalized as 
\begin{equation}
\overline{\textrm{FA}}_l = \frac{\textrm{FA}_l-\mu}{\sqrt{\delta^2+ \epsilon}},
\end{equation}
where $\epsilon >0$ is a small positive constant, and the parameters $\mu$ and $\delta$ are the mean and the standard deviation of $\textrm{FA}_l$, respectively. 

Then, the normalized features are subjected to element-wise summation and \textit{sigmoid} activation operation \cite{Giovanni2017} to generate the final channel attention map $\textrm{WA}_l$ as 
\begin{equation}
    \textrm{WA}_l = \textit{sigmoid}\left( \overline{\textrm{FA}}_l +  \overline{\textrm{FM}}_l\right),
\end{equation}
where $\overline{\textrm{FM}}_l$ is the normalized features of maxpooling layer.

Note that the generated channel attention weights vary depending on the input data (i.e., images), as shown in Fig. \ref{fig:attention} in the experimental results. To measure the importance of the channels, we use the statistical information found by element-wise averaging the weight in the channel attention map for all training data. The importance of channel $c$ for the intermediate tensor of CL $l$, $\omega_l^{c} \in \textrm{WA}_l$, can be calculated as
\begin{equation}
    \begin{array}{*{20}{c}}
        \omega_l^{c} = \frac{1}{|\mathcal{Z} |}\sum \limits_{z \in \mathcal{Z} } \omega_l^{c},&  1 \leq c \leq C_l,
    \end{array}
    \label{eq.channel}
\end{equation}
where $\mathcal{Z} $ is the training dataset with size $|\mathcal{Z} |$.

Finally, according to the compression ratio $m \in \mathcal{M}$, the original output tensor of CL $l$ can be compressed by pruning the less important channels and, thus, it outputs the compressed intermediate tensor $\overline{\mathcal{X}}_l \in \mathbb{R}^{\overline{C}_l \times H_l \times W_l}$. Note that the number of channels of the compressed tensor is $\overline{C}_l = C_l/m$.
\begin{figure*}[!t]

    \setcounter{equation}{15}
    \begin{equation}
        T_{u,s,k}^{\textit{arr}} = \left\{ {\begin{array}{*{20}{c}}
            t_{u,k}^{\textit{cmp}} + t_{u,s,k}^{\textit{com}},  &{k = 1},\\
            \max \left(T_{u,s'\in\mathcal{N}, k-1}^{\textit{arr}}, \left(k-1\right)\tau + t_{u,k}^{\textit{cmp}}\right) +  t_{u,s,k}^{\textit{com}}, &{k \neq 1}. 
        \end{array}} \right.
        \label{eq10}
        \end{equation}
        \begin{equation}
            t_{u,s,k}^{\textit{que}}  = \mathop {\arg \max }\limits_{k' \in \mathcal{K},u' \in \mathcal{U}} \left\{ {{\bf{1}}\left(T_{u,s,k}^{\textit{arr}}-T_{u',s, k'}^{\textit{arr}} \right) \cdot \left(\underbrace{T_{u',s, k'}^{\textit{arr}} + t_{u',s, k'}^{\textit{que}}+t_{u',s, k'}^{\textit{cmp}}}_{\text{Completion time instant of } u'{\text{'s task}}} - T_{u,s,k}^{\textit{arr}} \right)} \right\}.
            \label{eq11}
        \end{equation}
        \hrulefill

\end{figure*}
\subsection{Feature recovery module}
Since the computational operation of CNN is essentially a series of linear and nonlinear transformations, some redundant features can be obtained from other features by performing inexpensive nonlinear transformation operations \cite{Han_2020_CVPR}. Motivated by this, we design a lightweight CNN-based FR module to recover the intermediate tensor of CL $l$ from the received compressed information. 

As entropy coding is lossless, the entropy decoding yields the original compressed intermediate tensor $\overline{\mathcal{X}}_l$. Therefore, we just need to generate the channels pruned by the CA module using $\overline{\mathcal{X}_l}$, thereby rebuilding the intermediate tensor $\mathcal{X}_l$, as shown in Fig. \ref{fig:AECNN}(c). Unlike previous work \cite{Han_2020_CVPR}, we use all the channels of the received tensor to generate each pruned channel, which allows better learning and recovery of the representation of the channels that are pruned. 

To illustrate the feature recovery module, we use a function $f_R\left(\cdot\right)$ to represent the computation operation of learning the $c \, $th channel pruned by the CA module. Thus, the recovered $c \, $th channel can be denoted as
\setcounter{equation}{14}
\begin{equation}    
    \begin{array}{*{20}{c}}
            \mathcal{X}_l^c = f_R \left(\overline{\mathcal{X}}_l\right), & c \in \{\overline{C}_l +1, \cdots, C_l \} , 
    \end{array}
\end{equation}
where the recovered $C_l - \overline{C}_l$ channels will be concatenated to the received tensor $\overline{\mathcal{X}}_l$ as the input for CL $l+1$.

\section{Problem Formulation}\label{problem}
In this section, we first present the completion time a CNN inference task. Then, we formulate an optimization problem to maximize the average inference accuracy and throughput of inference tasks in a long-term perspective, while considering the energy consumption and transmission power constraints of an IoT device.

\subsection{Task Completion Time and Energy Consumption}
We assume that CNN inference tasks are processed on a first-come-first-served basis. In other words, an IoT device or ES can start processing a newly arrived task only after it has finished processing all previous arrivals. Let us consider the scenario where IoT device $u$ generates a task at timeslot $k$, which is offloaded to ES $s$. The IoT device can only initiate the transmission of this task after completing the transmission of its previous tasks. Neglecting propagation time, we assume the task generated by IoT device $u$ at timeslot $k$ arrives at ES $s$ at time instant $T_{u,s,k}^{\textit{arr}}$, and express it as (\ref{eq10}). Additionally, 
we use a function ${\bf{1}}\left(\cdot \right)$ to indicate the task of another IoT device $u'$ arrives at ES $s$ before IoT device $u$'s task, and calculate the queuing delay of IoT device $u$, $t_{u,s,k}^{\textit{que}}$, as  (\ref{eq11}).

There are three steps to complete computation offloading. First, an IoT device sends an inference task to an ES over wireless uplink, and then ES performs inference. Finally, ES sends the inference result back to the IoT device via downlink. Therefore, the completion time of an inference task includes communication time, queuing delay and computation time, which is denoted as
\setcounter{equation}{17}
\begin{equation}
    t_{u,s,k} = t_{u,k}^{\textit{cmp}} + t_{u,s,k}^{\textit{com}} + t_{u,s,k}^{\textit{que}}+ t_{u,s,k}^{\textit{cmp}}.
\end{equation}

Regarding the energy consumption of an IoT device, it encompasses both computation energy and communication energy. As such, we can calculate the total energy consumption of IoT device $u$ as 

\begin{equation}
    E_{u,s,k} = E_{u,k,l}^{\textit{cmp}} + E_{u,s,k}^{\textit{com}}.
\end{equation}

\subsection{Objective}
The goal of CNN inference offloading is to maximize the average inference accuracy and throughput of inference tasks in a long term perspective by designing a reasonable
computation offloading policy and resource scheduling policy. To achieve this goal, we define a reward function, $\varUpsilon \left(\mathcal{G}_k, \mathcal{I}_k, \mathcal{A}_k\right)$ to denote the achieved reward at timeslot $k$, as below
\begin{equation}
    \varUpsilon \left(\mathcal{G}_k, \mathcal{I}_k, \mathcal{A}_k   \right) =\sum\limits_{u \in \mathcal{U}} \sum\limits_{s \in \mathcal{S}} \sum\limits_{m \in \mathcal{M}}  {\eta _{u,m}^k \psi \left(t_{u,s,k}\right)}, 
    \label{eq:reward}
\end{equation}
where $\mathcal{A}_k \triangleq \left\{ \alpha_{u,s}^k \beta_{u,l}^k \gamma_{u,m}^k| u \in \mathcal{U}, s \in \mathcal{S}, l \in \mathcal{L}, m \in \mathcal{M} \right \}$ is the offloading decision determining the computing mode, the matching between IoT devices and ES, and the compression ratio of transmitted intermediate tensor; the function $\psi(x)$ introduces a penalty mechanism for tasks that exceed their deadlines. This penalty function mediates the trade-offs among communication efficiency, computation capability and inference accuracy. The definition of $\psi(x)$ is introduced in Theorem \ref*{theorem1} and the proof is detailed in Appendix \ref{proof}. 
\begin{theorem} \label{theorem1}
    Let $\psi(x) \triangleq 2\left(1 - {\textit{sigmoid}}\left(\frac{5x}{\sigma_u^k}\right)\right)$. For any completion time $t_{u,s,k}$ and latency requirement $\sigma_u^k$, we have:
    \begin{enumerate}
        \item As $t_{u,s,k}$ approaches $\sigma_u^k$, $\psi(t_{u,s,k}) \rightarrow 0$.
        \item As $t_{u,s,k}$ approaches 0, $\psi(t_{u,s,k}) \rightarrow 1$.
    \end{enumerate}
\end{theorem}

Accordingly, we express 
the average achieved reward function over a period as below:
\begin{equation}
    \textit{Q}\left(\mathcal{K}, \mathcal{G},\mathcal{I}, \mathcal{A}\right) =  \frac{1}{K} \sum\limits_{k \in \mathcal{K}} \varUpsilon \left(\mathcal{G}_k,\mathcal{I}_k, \mathcal{A}_k\right).
\end{equation}

The optimization problem of maximizing the average accuracy and throughput of inference tasks over a period is based on the above reward function. It is a mixed integer programming non-convex problem that is difficult to solve with conventional algorithms. To address this issue, we decouple it into two subproblems, $\mathcal{P}_1$ is the offloading strategy:
\begin{align}
    \mathcal{P}_1: & \max\limits_{\mathcal{K}, \mathcal{A}} \;  \textit{Q}\left(\mathcal{K}, \mathcal{G},\mathcal{I}, \mathcal{A}\right) \label{p1} \\
    &  \;\; \;\textrm{s.t.}  \;\; \alpha_{u,l}^k \in \{0,1\},  \forall u \in \mathcal{U} \; \textrm{and} \; \forall l \in \mathcal{L}, \tag{\ref{p1}{a}} \label{p1a} \\
    &  \; \;\;\; \;\;\;\;\; \beta_{u,s}^k \in \{0,1\}, \forall u \in \mathcal{U} \; \textrm{and} \; \forall s \in \mathcal{S}, \tag{\ref{p1}{b}} \label{p1b} \\
    & \; \;\;\; \;\;\;\;\; \gamma_{u,m}^k \in \left\{0,1 \right\}, \forall u \in \mathcal{U} \; \textrm{and} \; \forall m \in \mathcal{M} \tag{\ref{p1}{c}} \label{p1c}
\end{align}

Once the optimal computation offloading decision $\mathcal{A}^* = \{\mathcal{A}_1^*, \mathcal{A}_2^*,\cdots, \mathcal{A}_K^*\}$ is determined, the optimization problem is simplified to a convex optimization problem $\mathcal{P}_2$
to optimize the resource allocation:
\begin{align}
    \mathcal{P}_2: & \max\limits_{ \mathcal{K}}   \;\textit{Q}\left(\mathcal{K}, \mathcal{G},\mathcal{I}, \mathcal{A}^*\right) \label{p2} \\
    & \;\; \;\textrm{s.t.}  \;\; t_{u,s,k} \leq \sigma_{u}^k, \forall u \in \mathcal{U} \; {\rm{and}} \; s \in \mathcal{S}, \tag{\ref{p2}{a}} \label{p2a} \\
    &  \;\;\;\; \;\;\;\;\; p_{u,s}^k \leq P_{u}, \forall u \in \mathcal{U} \; \textrm{and} \; \forall s \in \mathcal{S}, 
    \tag{\ref{p2}{b}} \label{p2b} \\
    &  \; \;\;\; \;\;\;\;\;E_{u,s,k} \leq E_{u,k,L}^{\textit{cmp}}, \forall u \in \mathcal{U} \; \textrm{and} \; \forall s \in \mathcal{S}. \tag{\ref{p2}{c}} \label{p2c}
 \end{align}
Note that the constraint (\ref{p2c}) ensures that offloading is more energy-efficient than performing the computation locally.

\begin{figure*}[t]
    \centering
    \includegraphics[width= \textwidth]{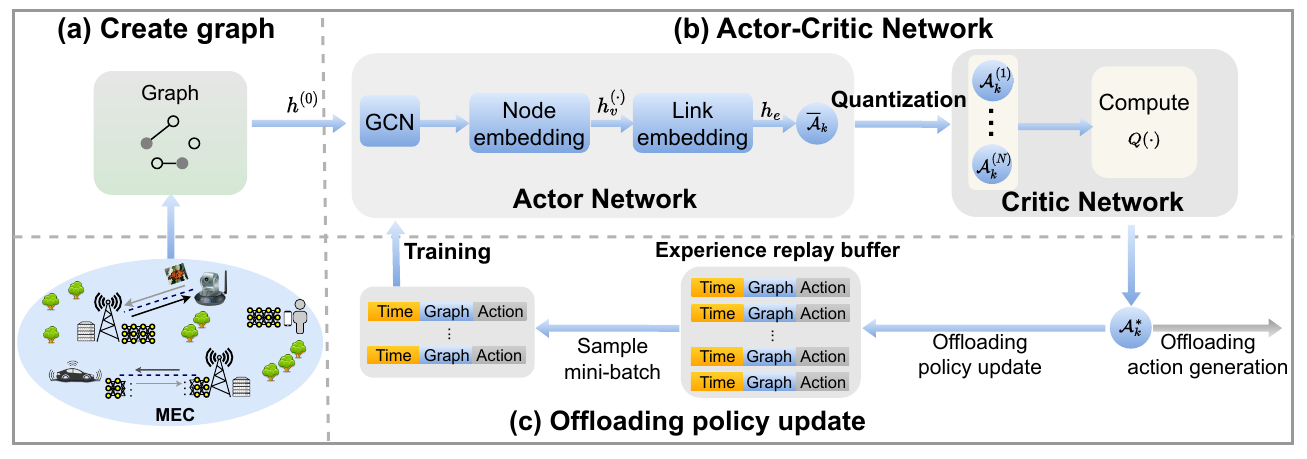}
    \caption{Framework of graph reinforcement learning-based AECNN}
    \label{fig:graph}
\end{figure*}
\section{Graph reinforcement learning-based AECNN}\label{method}
In this section, we present the framework of our proposed GRL-AECNN to address the optimization problem described in $\mathcal{P}_1$. Subsequently, we provide a comprehensive overview of GRL-AECNN and outline the training strategy.
\subsection{GRL-AECNN framework}
In dynamic MEC networks, the data exhibits a graph-like structure rather than a regular Euclidean format. To effectively handle such graph data, we propose GRL-AECNN by applying GCN \cite{Mathias2016} to analyze the characteristics of graph data through message passing and aggregation between nodes, as shown in Fig. \ref{fig:graph}. By learning the aggregation method based on the relationships between nodes, GCNs can effectively process and understand the graph-like characteristics of the data.
Additionally, GRL-AECNN can automatically filter out messages from disconnected nodes through graph data updates, which obviates the need for retraining the aggregation function when facing a new MEC network topology. Consequently, the GRL-AECNN exhibits robust adaptability in handling changes within the dynamic MEC network structure, without necessitating extensive reconfiguration.

In the proposed GRL-AECNN framework, an {\textit{actor-critic}} network is used to generate offloading decisions and update offloading policies. The actor network is responsible for predicting actions; the critic network quantifies the prediction and generates offloading decisions; and the experience replay buffer stores historical experiences and samples mini-batch training data to train the GCN. Since each task can only be split at one layer $l$, compressed by one compression ratio $m$ then offloaded to one ES $s$, the three-step task offload decision for device $u$'s task can be merged into one step, i.e., an IoT device has $(L+1)MS$ options to perform its task. In GRL-AECNN, we model the structure information of the MEC scenario at timeslot $k $ as graph data $\Gamma_k = \left(\mathcal{V}_k,\mathcal{E}_k\right)$, where $M$ devices and $(L+1)MS$ options are represented by the graph vertices $\mathcal{V}_k$, and each IoT device and option is connected by a directed edge $e \in \mathcal{E}_k$.
\subsection{Actor network}
In the actor network, we represent the feature of the $i$th GCN layer as $h^{\left(i\right)}= \left\{ h_v^{\left(i\right)} | v \in \mathcal{V}_k \right \}$. Specially, we parse the 
MEC state $\Gamma_k$ as the initial input data $h^{\left(0\right)}$ for GCN. GCN uses multiple graph convolutional layers to aggregate the neighborhood information. For each node $v \in \mathcal{V}_k$, we define the neighborhood information aggregation
process as follows,
\begin{equation}
    h_v^{\left(i+1\right)}= {\textit{Relu}}\left(\varpi ^{\left(i+1\right)} \mathbf{C} \left(h_v^{\left(i\right)}, \mathbf{A}  ^{\left(i\right)}{\left(h_{v'}^{\left(i\right)} \right)}\right)\right), v' \in \varrho_v
    \label{eq:aggregation}
\end{equation}
where $\varpi  ^{\left(i+1\right)}$ is the weight parameters, $\mathbf{A} ^{\left(i\right)}\left(\cdot \right)$ is the aggregation function, $\varrho_v$ is the set of node $v$'s neighbors, $\mathbf{C} \left(\cdot \right)$ is a concatenate operation, $\textit{Relu}\left(\cdot \right)$ is a non-linear function \cite{Giovanni2017}. 

The system can acquire the information of tasks and the status of ESs by aggregating the information in the second-order neighborhood of nodes. For example, IoT device $u$ can grasp the information of its second-order neighborhood (other IoT devices connected to ES $s$) through its first-order neighborhood
ES $s$; ES $s$ can acquire the status of its second-order neighborhood (other ESs) through its first-order neighborhood
(IoT devices connected to ES $s$). Therefore, we use two GCN layers in GRL-AECNN, i.e., $i \in \left\{0,1\right\}$ in (\ref{eq:aggregation}).

Once the information aggregation of nodes is finished, the next step is to 
obtain the feature representation of edge $e \in \mathcal{E}_k$, $h_e$, through concatenating the features of its source node $v' \in \mathcal{V}_k$ and destination node $v'' \in \mathcal{V}_k$. This process is outlined as
\begin{equation}
    h_e = \mathbf{C}  \left(h_{v'}^{\left(2\right)}, h_{v''}^{\left(2\right)} \right).
\end{equation}

Then, we can classify the edges to get the relaxed offloading action $ \zeta_k = \left\{{\overline a}_{k,e}|{\overline a}_{k,e}=  \textbf{F} \left( {h_e} \right), e \in \mathcal{E}_k \right \}$, by the function 
\begin{equation}
    \textbf{F}\left(h_e\right) = {\textit{sigmoid}}\left({\textit{MLP}}_2\left({\textit{Relu}}\left({\textit{MLP}}_1\left(h_e\right)\right)\right)\right),
    \label{eq.action1}
\end{equation}
where ${\textit{MLP}}_1$ and ${\textit{MLP}}_2$ are multi-layer perceptions to extract the feature of edge $e$. We use $\textit{sigmoid}\left(\cdot \right)$ function to make the relaxed offloading action satisfy $0 < {\overline a}_{k,e} < 1$ \cite{Giovanni2017}. 
\subsection{Critic network}
In critic network, we first use the order-preserving method in DROO \cite{Huang2020} to quantify the relaxed offloading action $\zeta_k$ and generate $N=ULMS$ candidate binary offloading decisions $\mathbb{A}_k = \left\{ \overline{\zeta}_k^{\left(1\right)},\overline{\zeta}_k^{\left(2\right)},\cdots,\overline{\zeta}_k^{\left(N\right)} \right \}$, where $\overline{\zeta}_k^{\left(n\right)} = \left\{ a_{k,e}^{\left(n\right)}|a_{k,e}^{\left(n\right)} \in \left\{0,1\right\}, e \in \mathcal{E}_k \right \}$.

Recall that each candidate offloading action $\overline{\zeta}_k^{\left(n\right)}$ can achieve reward by solving (\ref{eq:reward}). Therefore, the optimal offloading action at $k\rm{th}$ timeslot can be generated as
\begin{equation}
    \mathcal{A}_k^* =\arg \mathop {\max }\limits_  {\overline{\zeta}_k^{\left(n\right)} \in \mathbb{A}_k} \textit{Q}\left(\mathcal{K}, \mathcal{G},\mathcal{I}, \mathcal{A}\right).
    \label{eq.action2}
\end{equation}
\subsection{Complexity Analysis and Training Strategy}
\subsubsection{Complexity Analysis}
At timeslot $k$, the computational complexity associated with the offloading decision $\mathcal{A}_k$ is represented as $ULMS$. However, numerous IoT devices and multiple candidate splitting points of a CNN model may cause substantial complexity. In fact, not all 
the candidate splitting points are meaningful for decision-making. As described in Section \ref{measurement}, with the same communication overhead, splitting the CNN model at the deeper split points may result in lower inference accuracy while increasing the computation overhead on the IoT device. As such, we first train the proposed AECNN and select the meaningful splitting points that are used to train the GRL-AECNN thereby performing the offloading decision-making. 

\subsubsection{Training of AECNN}The proposed AECNN architecture can be trained in an end-to-end manner. However, this may result in very slow convergence. Therefore, we use a step-by-step training approach to train our proposed AECNN. We first insert the designed CA module into the original CNN model and then train the resulting neural network to figure out the importance of the channels. Based on the statistic of channels' importance, for a given compression ratio $m$, $C_l(1-\frac{1}{m})$ channels with the lowest importance are identified as prunable. 
Then, we remove the inserted CA module and prune the original CNN model by removing the identified channels and the corresponding filters. Next, we fine-tune the pruned CNN model to recover the accuracy loss caused by the model pruning. Finally, we insert the designed FR module into the pruned CNN model and fine-tune the resulting CNN model to improve the inference accuracy. Throughout the training process, we do not consider the entropy encoding and decoding modules, because this lossless compression does not cause any accuracy loss. The detailed training process is described in Algorithm \ref{alg1}.
\begin{algorithm}[t]
\renewcommand{\algorithmicrequire}{\textbf{Input:}}
\renewcommand{\algorithmicensure}{\textbf{Output:}}
    \caption{Training strategy of AECNN}
    \label{alg1}
    \begin{algorithmic}[1]
    \REQUIRE CNN model $\Omega$, the set of CLs $\mathcal{L}$, the set of compression ratio $\mathcal{M}$, the set of training data $\mathcal{Z} $.
    \ENSURE  The set of AE-enhanced CNN models $\Omega = \left\{\Omega_1^2, \cdots, \Omega_1^M, \cdots,\Omega_{L-1}^M \right\}$.
    \FOR{$l=1$ to $L-1$}
        \STATE Insert CA module after the splitting point $l$.
        \STATE Train the resulting network on training data $\mathcal{Z}$.
        \STATE Calculate the importance of each channel using (\ref{eq.channel}).
        \STATE Sort the importance of all the channels.
        \STATE Remove the inserted CA module.
        \FOR{$m=2$ to $M$}
        \STATE Compress CL $l$ by pruning $C_l \left(1- \frac{1}{m}\right)$ less important channels. 
        \STATE Fine-tune the pruned CNN model.
        \STATE Insert FR module into the pruned CNN model before CL $l+1$, then fine-tune the resulting neural network to get $\Omega_l^m$. 
        \ENDFOR
    \ENDFOR
    \RETURN $\Omega$
    \end{algorithmic}
\end{algorithm}
\begin{algorithm}[t]
\renewcommand{\algorithmicrequire}{\textbf{Input:}}
\renewcommand{\algorithmicensure}{\textbf{Output:}}
    \caption{GRL-AECNN for offloading decision-making}
    \label{alg2}
    \begin{algorithmic}[1]
    \REQUIRE Input MEC state $\Gamma_k, \forall k \in \mathcal{K}$, training interval $\omega$.
    \ENSURE Output offloading decision $\mathcal{A}_k^*$.
    \FOR{$k=1$ to $K$}
        \STATE Generate the relaxed offloading action $\overline{\zeta}_k$ in (\ref{eq.action1}).
        \STATE Quantify $\overline{\zeta}_k$ into $N$ binary actions $\mathbb{A}_k$.
        \STATE Select the optimal offloading action $\mathcal{ A}_k^*$ using (\ref{eq.action2}).
        \STATE Update the experience replay buffer by adding $\left(\Gamma_k,\mathcal{ A}_k^*\right)$.
        \IF{$k$ \rm{mod} $\omega$ = 0}
        \STATE Randomly sample a mini-batch of training data $\Delta_k$ from the buffer.
        \STATE Train GCN and update the parameters using (\ref{eq.action3}).
        \ENDIF
    \ENDFOR
    \RETURN $\mathcal{A}_k^*$
    \end{algorithmic}
\end{algorithm}
\subsubsection{Offloading policy update}
We use the experience replay buffer technique to train the GCN using the stored data samples $\left(k,\Gamma_k,\mathcal{A}_k^*\right)$, as shown in Fig. \ref{fig:graph}. At timeslot $k$, we randomly select a mini-batch of training data $\Delta_k = (\Delta_k^\mathcal{T}, \Delta_k^\Gamma, \Delta_k^{\mathcal{A}^*})$ from the memory to update the parameters of GCN and reduce the averaged cross-entropy loss \cite{Huang2020}, as
\begin{equation}
    \begin{split}
        \xi \left(\Delta_k\right) =  -\frac{1}{|\Delta _k|}\sum\limits_{k' \in {\Delta_k^\mathcal{T}}} & 
        \left(1-\mathcal{A}_{k'}^*\right) \log \left(1-{\textit{f}}_I \left( \mathcal{E}_{k'} \right)\right)  \\
        & + \mathcal{A}_{k'}^*\log {\textit{f}}_I \left( \mathcal{E}_{k'} \right),
    \end{split}
    \label{eq.action3}
\end{equation}
where $|\Delta _k|$ is the size of the mini-batch training data, $\Delta_k^\mathcal{T}$ is the set of timeslots, $\Delta_k^\Gamma$ is the set of graphs, and $\Delta_k^{\mathcal{A}^*}$ is the set of actions. The detailed process of GRL-AECNN is described in Algorithm \ref{alg2}.

\section{Performance Evaluation}\label{Performance}
\subsection{Experimental Setup}
We consider an MEC network comprising of $S=2$ ESs (RTX 2080TI GPU) located at $ \left [ \left(30\textrm{m}, 30\textrm{m}\right), \left(90\textrm{m}, 30\textrm{m}\right) \right]$, and $U=14$ IoT devices (Raspberry pi 4B) randomly
distributed in the $\left[0, 120 \right]\times\left[0, 60 \right] \textrm{m}^2$ region. 
The bandwidth for the 2.4 GHz WiFi connection between the IoT device and the respective ES is set at $B_{u,s}^k = 20$ MHz, the noise power spectral density is $n_0 = -174$ dBm/Hz and the maximal transmission power of each IoT device is limited to $P_u = 20$ dBm. Similar to\cite{Huang2020}, we consider the free-space propagation model and express the average channel gain as $\overline{g}_{u,s}^k =g_a(\frac{3 \times 10^8}{4 \pi f_c \vartheta_{u,s}^k})^{d_e}$, where $f_c$ is the WiFi frequency, $g_a = 2 $ is the antenna gain, $d_e = 2.8 $ is the path loss exponent, and $\vartheta_{u,s}^k$ represents the distance between IoT device $u$ and ES $s$, measured in meters. The wireless channel gain $g_{u,s}^k$ can be expressed as $g_{u,s}^k=\overline{g}_{u,s}^k g_r$,  where the Rayleigh small-scale fading coefficient follows $g_r \sim \mathcal{CN}(0,\textbf{I})$.
Without loss of generality, we assume that channel gains remain consistent within a single timeslot and exhibit independent variability from one timeslot to another.

We consider the classification task of Caltech-101 dataset \cite{caltech101}, consisting of approximately 9,000 images categorized into 101 classes. Each category comprises roughly 40 to 800 images with resolutions ranging from $200\times 200$ to $300 \times 300$ pixels. The diversity in resolutions aligns well with the variability typically encountered in IoT applications. We assume task sizes $d_{m,n}^k$ ranging between 20 KBytes and 100 KBytes, 
and use the popular ResNet-50 \cite{He_2016_CVPR} for image classification. To perform the co-inference, we split ResNet-50 at different splitting points and compress the intermediate tensor with different compression ratios $\mathcal{M} = \left\{1, 2,4,8,16,32,64\right\}$. Since ResNet-50 introduces a branching structure with residual blocks instead of a sequential structure, the first CL and each residual block are considered as the candidate splitting points.

We use PyTorch to implement GRL-AECNN with the following training parameters: the hidden neurons of two GCN layers are 128 and 64, the learning rate is initialized as 0.001, the experience replay buffer size is 128, the mini-batch size $|\Delta _k| = 64$;  the training interval $\omega=10$, and the optimizer for the loss function $\xi \left(\Delta_k\right)$ is the Adam function \cite{004}. To validate the effectiveness of GRL-AECNN, we conduct a comprehensive evaluation comparing semantic compression performance and CNN inference offloading efficiency. We first compare the performance of AECNN with existing state-of-the-art semantic compression methods, BottleNet++ \cite{BottleNet++} and DeepJSCC \cite{Jankowski2021JSAC}. In both BottleNet++ and DeepJSCC, the intermediate tensor is encoded using a CNN-based encoder with dimension adjustment at the final FC layer. Subsequently, we compare the performance of GRL-AECNN with the state-of-the-art CNN inference offloading method, DROO \cite{Huang2020}. Our comparative analysis involves the following three methods: 
\begin{itemize}
    \item DROO-AECNN: DROO \cite{Huang2020} enhanced AECNN; 
    \item  GRL-BottleNet++: GRL enhanced BottleNet++ \cite{BottleNet++}; 
    \item GRL-DeepJSCC: GRL enhanced DeepJSCC \cite{Jankowski2021JSAC}.
\end{itemize}

\begin{figure}[t]
    \centering
    \includegraphics[width=0.45 \textwidth]{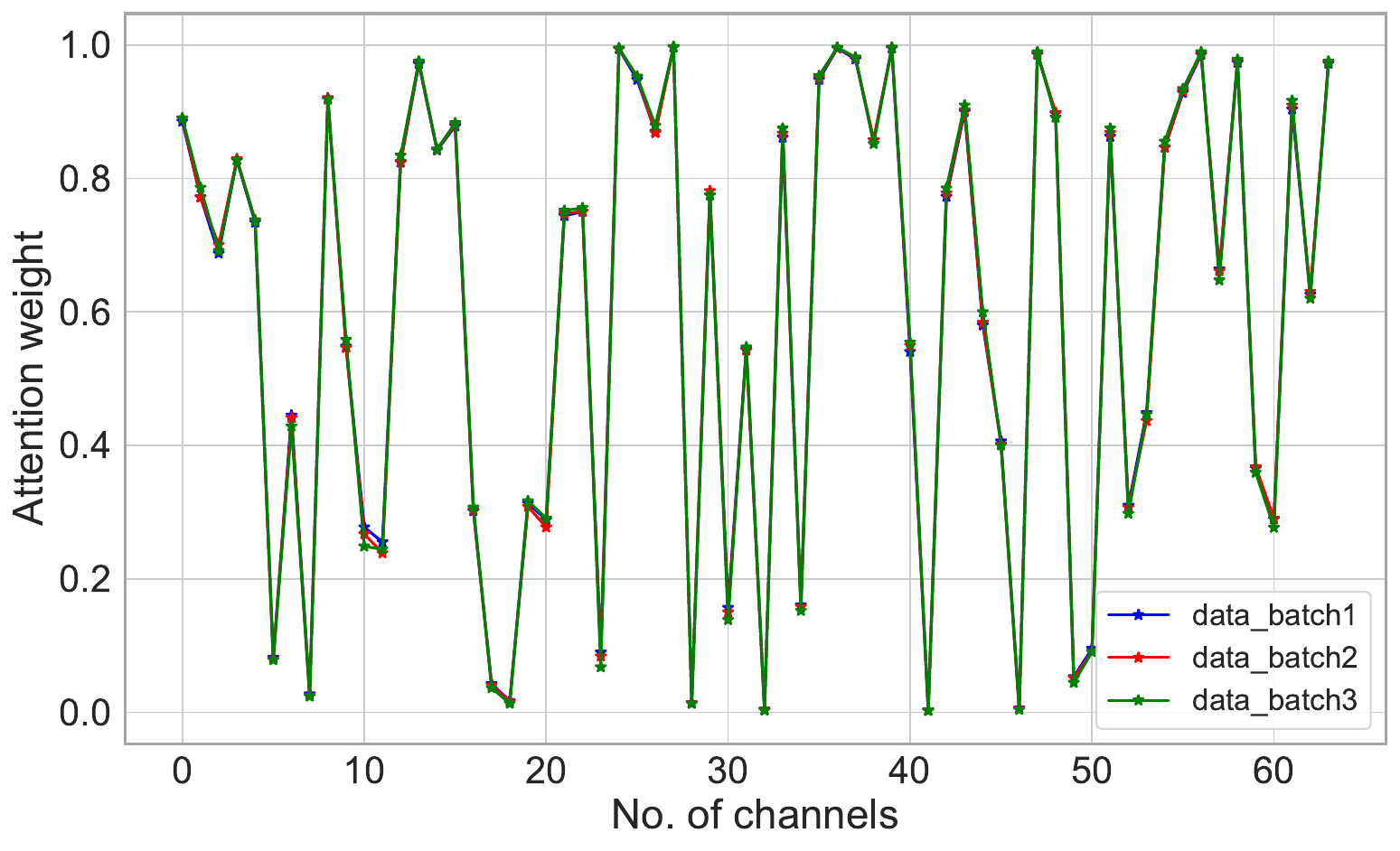}
    \caption{Attention weights of splitting point $l=1$.}
    \label{fig:attention}
\end{figure}

\renewcommand\arraystretch{1.3}
\begin{table*}
    \centering
      \caption{Inference accuracy under different compression ratios of intermediate tensor and entropy encoding (measured on Raspberry pi 4B) and decoding (measured on RTX 2080TI) time}
    \label{tab:BitErrors}
    \centering
    \resizebox{\textwidth}{!}{
      \begin{tabular}{ccccccccccccccccc}
        \toprule
     \makecell[c]{Splitting \\point} &$C_l \times H_l \times W_l$ & $m$ & BottleNet++ (\%) & JSCC (\%) & CA\_Pruned (\%) &AECNN (\%) &$C'_l$  &Entropy (bit) & $t_{u,l,m}^{\textit{enc}}$ (ms)  &$t_{s,l,m}^{\textit{dec}}$ \\ 
        \midrule
     & & $2\times$ & $92.19(\pm0.22)$ & $93.10(\pm0.15)$ &$95.58(\pm0.24)$  &$95.62(\pm0.20)$ &$32$  &$11.13(\pm0.18)$ &4.53 &3.03\\ 
     & & $4\times$ & $\underline{\mathbf{91.66(\pm0.27)}}$ & $\underline{\mathbf{92.06(\pm0.21)}}$ &$95.05(\pm0.26)$  &$\underline{\mathbf{95.30(\pm0.21)}}$ &$16$  &$10.48(\pm0.16)$ &3.25 &2.41\\ 
     $l=1$ &$64\times 56 \times 56$ & $8\times$ & $91.49(\pm0.36)$ & $91.83(\pm0.23)$ &$94.57(\pm0.11)$  &$94.68(\pm0.17)$ &$8$  &$9.86(\pm0.21)$ &1.94 &1.52\\ 
     && $16\times$ & $90.92(\pm0.23)$ & $91.29(\pm0.19)$ &$93.89(\pm0.54)$  &$93.98(\pm0.17)$ &$4$  &$9.13(\pm0.27)$ &1.76 &1.37\\ 
     & & $32\times$ & $89.76(\pm0.18)$ & $90.76(\pm0.18)$ &$92.69(\pm0.65)$  &$92.70(\pm0.31)$ &$2$  &$8.62(\pm0.12)$ &1.03 &0.74\\ 
    & & $\underline{\mathbf{64\times}}$ & $88.54(\pm0.44)$ & $89.52(\pm0.29)$ &$91.10(\pm0.53)$  &$\underline{\mathbf{91.47(\pm0.25)}}$ &$1$  &$\underline{\mathbf{7.73(\pm0.32)}}$ &0.62 &0.39\\ 
     & & $2\times$ & $93.28(\pm0.30)$ & $94.39(\pm0.27)$ &$95.43(\pm0.19)$  &$95.48(\pm0.33)$ &$128$  &$12.16(\pm0.27)$ &7.89 &4.76 \\ 
     & & $4\times$ & $92.25(\pm0.17)$ & $93.85(\pm0.24)$ &$95.23(\pm0.25)$  &$95.24(\pm0.28)$ &$64$  &$11.51(\pm0.29)$ &5.07 &3.21\\ 
     $l=2$ &$256\times 56 \times 56$ & $8\times$ & $91.70(\pm0.39)$ & $92.91(\pm0.31)$ &$94.96(\pm0.14)$  &$95.05(\pm0.25)$ &$32$  &$10.99(\pm0.24)$ &3.51 &2.60\\ 
     && $16\times$ & $91.64(\pm0.31)$ & $92.21 (\pm0.19)$ &$94.80(\pm0.30)$  &$94.85(\pm0.16)$ &$16$  &$10.42(\pm0.21)$ &3.00 &2.35 \\ 
     & & $32\times$ & $90.86(\pm0.28)$ & $91.65(\pm0.21)$ &$94.61(\pm0.21)$  &$94.64(\pm0.19)$ &$8$  &$9.82(\pm0.18)$ &1.87 &1.46\\ 
    & & $64\times$ & $90.63(\pm0.22)$ & $91.04(\pm0.31)$ &$93.63(\pm0.28)$  &$93.78(\pm0.19)$ &$4$  &$9.19(\pm0.22)$ &1.90 &1.44\\ 
       \bottomrule
    \end{tabular}}
    \end{table*}
\subsection{Performance of GRL-AECNN} \label{measurement}
\subsubsection*{\bf{Measurements of attention weights in AECNN}}
To verify the robustness of the statistical method for calculating the importance of channels in AECNN, we use the same amount of input data from different batches to calculate the importance of each channel. We hereby take the first candidate point as an example and calculate the importance of each channel for the intermediate tensor by randomly sampling three batches of input data. As shown in Fig. \ref{fig:attention}, we can see that the overall trend of the channels' importance is essentially consistent across these three batches of data. This means that while the importance of individual channels might vary depending on the specific input data, the general ranking and trend of the channels' importance remains relatively stable, which demonstrates the feasibility of the statistical method we used for calculating the importance of channels.
\subsubsection*{\bf{Inference accuracy of AECNN under various compression ratios}}
In partial offloading, the latency is mainly caused by the computation and communication time on the IoT device due to the limited resources and bandwidth. Therefore, we should split the CNN model as early as possible (near the input layer) to reduce the computation on the IoT device and compress the intermediate tensor as much as possible without compromising too much accuracy. In our experiment, we found that the computation time on the IoT device alone exceeds 100ms, if ResNet-50 is split at the third or later candidate points, which is not suitable for real-time inference. Therefore, we mainly consider the first and second candidate points. In our experiments, the inference accuracy of the original ResNet-50 on the test dataset is $95.84(\pm0.35) \%$.

Table \ref{tab:BitErrors} compares the inference accuracy of AECNN with that of BottleNet++ and DeepJSCC at different compression ratios of intermediate tensor, where `CA\_Pruned' signifies the pruned ResNet-50 without the FR module. We can see that AECNN improves the accuracy of CA\_Pruned ResNet-50, which demonstrates the effectiveness of the proposed FR module. In general, higher compression will result in more accuracy loss due to the lack of comprehensive presentation of the features. The experimental result shows that AECNN consistently outperforms BottleNet++ and DeepJSCC at different compression ratios. For example, when $l=1$ and $m=4$, AECNN improves the accuracy from 91.66\% and 92.06\% to 95.30\% in comparison with BottleNet++ and DeepJSCC, respectively. This improvement is attributed to AECNN's ability to extract informative features instead of directly compressing the intermediate tensor
as BottleNet++ and DeepJSCC do, thus avoiding the loss of semantic information. Moreover, AECNN achieves higher accuracy by splitting the model at the first splitting point than the second with the same communication overhead. For example, splitting the model at the first point, AECNN achieves an accuracy of 93.98\% when $m=16$, while that of the second point is 93.78\% when $m=64$. Note that in this case, the compressed data size at the first point is $4\times64\times56\times56/16$, which is equivalent to $4\times256\times56\times56/64$ at the second point. Therefore, choosing the first split point is a better option, in addition, it uses less computation time and can reduce
the overall task completion time.
At the first splitting point, AECNN can compress the intermediate tensor by more than $256 \times$ (i.e., $64 \times 32/7.73$) using channel pruning and entropy coding, with accuracy loss of only about 4\% (i.e., $95.84\%-91.47\%$). 
In subsequent experiments, we mainly focus on
the first splitting point in \textit{partial offloading} to alleviate the computation complexity of GRL-AECNN.

\subsubsection*{\bf{Convergence of GRL-AECNN}}
We first define the normalized reward for timeslot $k$ as
\begin{equation}
    {\overline {\varUpsilon }}  \left(\mathcal{G}_k, \mathcal{I}_k, \mathcal{A}_k   \right) = \varUpsilon   \left(\mathcal{G}_k, \mathcal{I}_k, \mathcal{A}_k   \right) / \varUpsilon   \left(\mathcal{G}_k, \mathcal{I}_k, \mathcal{A}'_k   \right) 
\end{equation}
where the action $ \mathcal{A}'_k$ is obtained by exhaustive searching.

\begin{figure*}[t]
    \centering
    \subfigure[Normalized reward]{
    \includegraphics[width=0.45\textwidth]{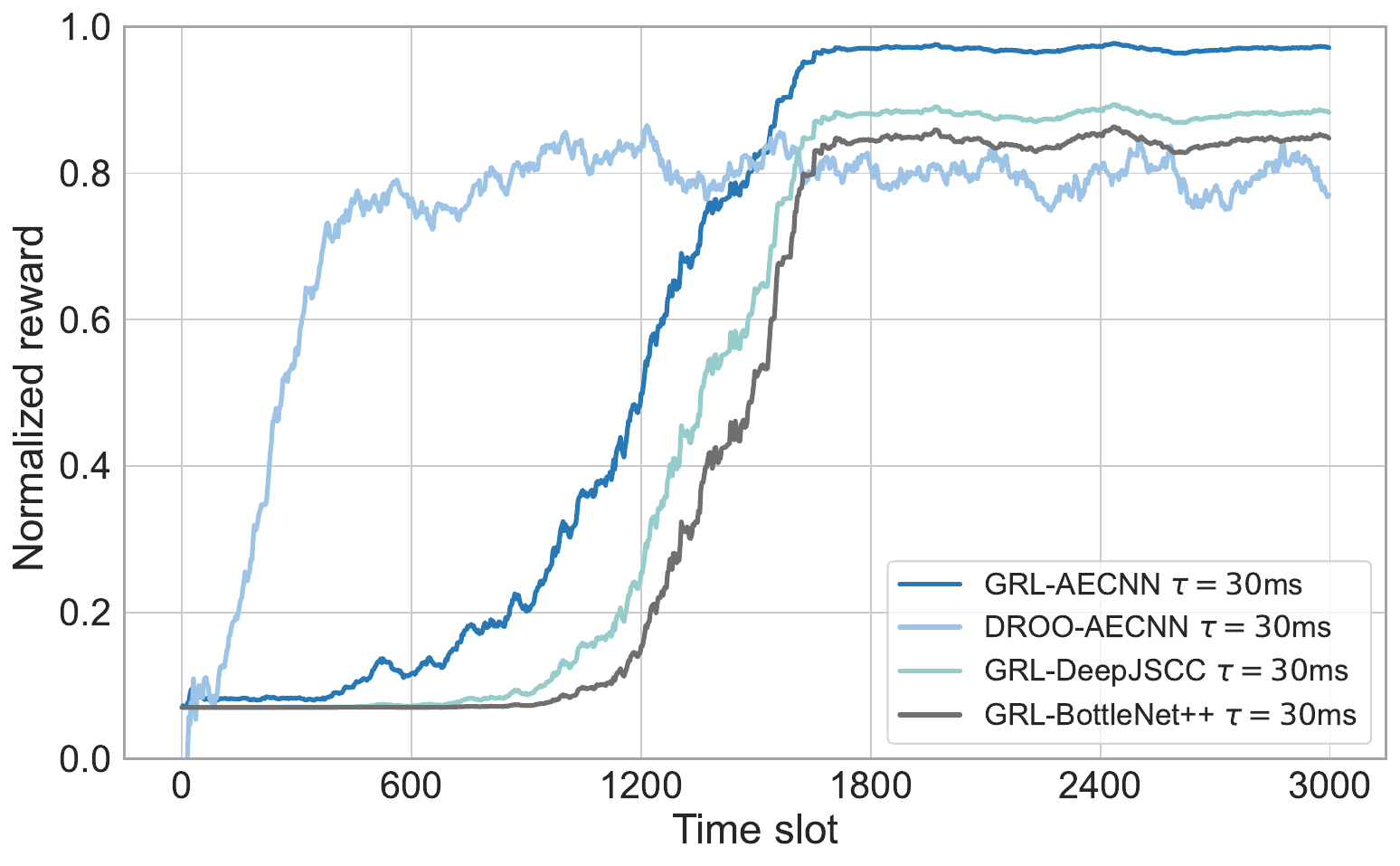}}
    \hspace{5 mm}
    \subfigure[Training loss]{
    \includegraphics[width=0.45\textwidth]{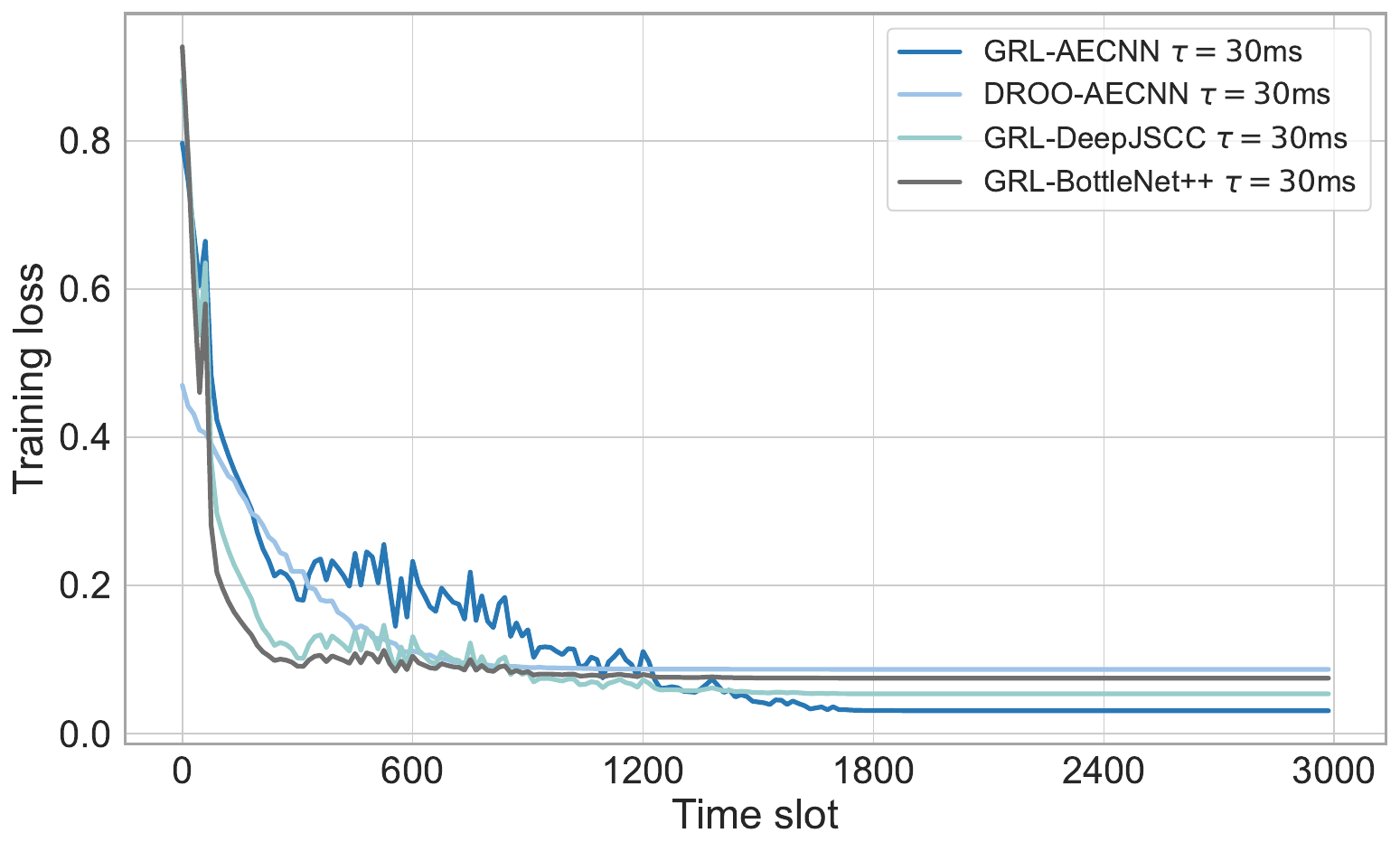}}
    \caption{Performance of convergence}
    \label{fig:convergence}
\end{figure*}

In Fig. \ref{fig:convergence}, we characterize the convergence performance by plotting the moving average of ${\overline {\varUpsilon}}$ over the most recent 50 timeslots, alongside the training loss. As the timeslot increases, the moving average of ${\overline {\varUpsilon}}$ and training loss show a gradual convergence towards the optimal solution. The occasional fluctuations are primarily attributed to the randomness of training data sampling. Specifically, as the timeslot increases, the moving average of the normalized reward ${\overline \varUpsilon}$ for GRL-AECNN consistently exceeds 0.95, while the training loss remains consistently below 0.03. This performance superiority is particularly evident when compared to the other three methods. This distinction can be attributed to GRL-AECNN's ability to make full use of the MEC states for making offloading decisions, thus providing an advantage over DROO-AECNN, which only considers the wireless channel state for decision-making. Compared to GRL-BottleNet++ and GRL-DeepJSCC, GRL-AECNN can effectively extract the semantic information to make better offloading decisions, resulting in superior performance. 
The robust convergence, coupled with the compelling performance metrics, reinforces the efficiency of GRL-AECNN in optimizing offloading decision-making.

\begin{figure*}[t]
    \centering
    \subfigure[Average accuracy]{
    \includegraphics[width=0.32\textwidth]{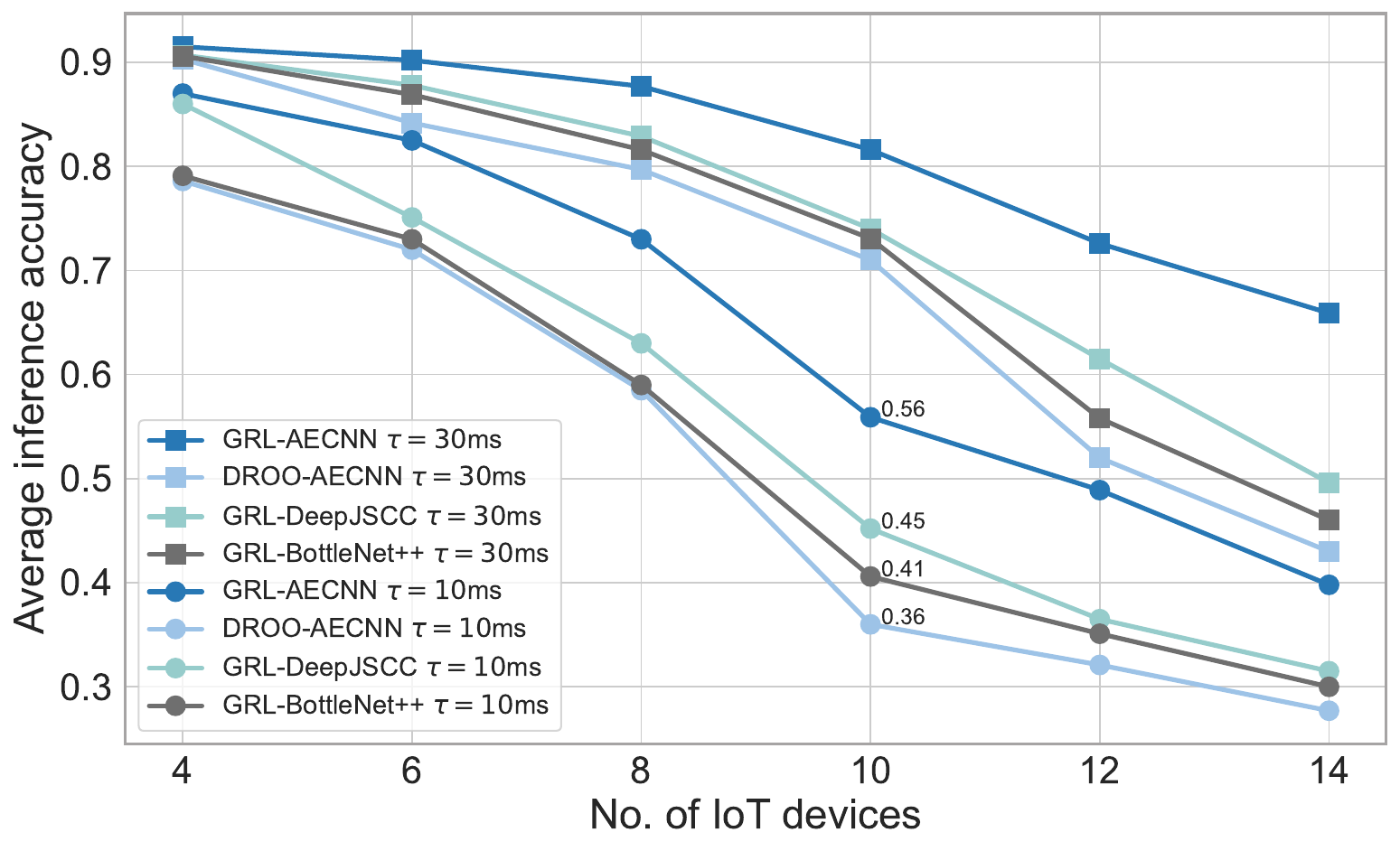}
    \label{full_acc}}
    \subfigure[Service successful probability]{
    \includegraphics[width=0.32\textwidth]{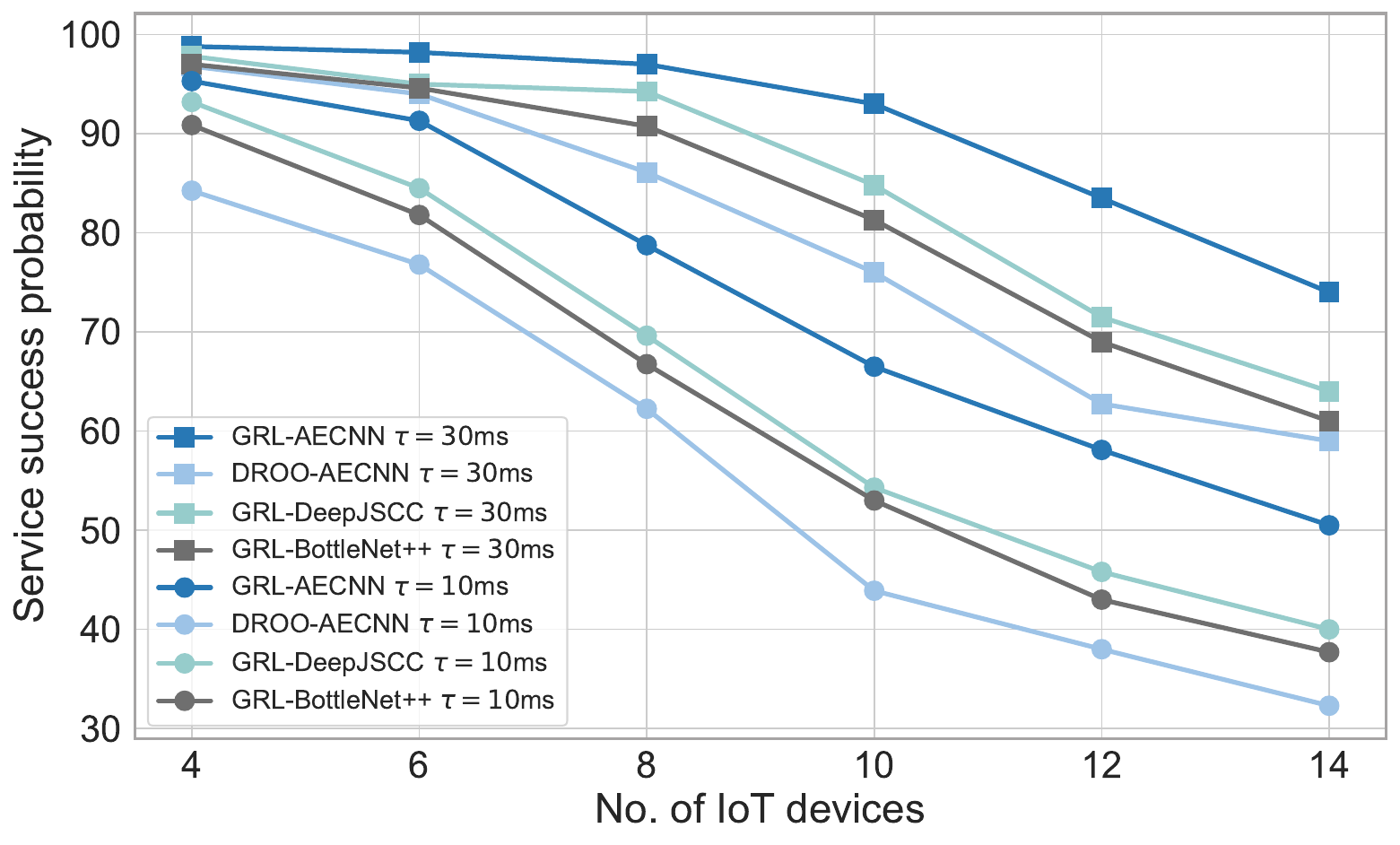}}
    \subfigure[Average throughput]{
    \includegraphics[width=0.32\textwidth]{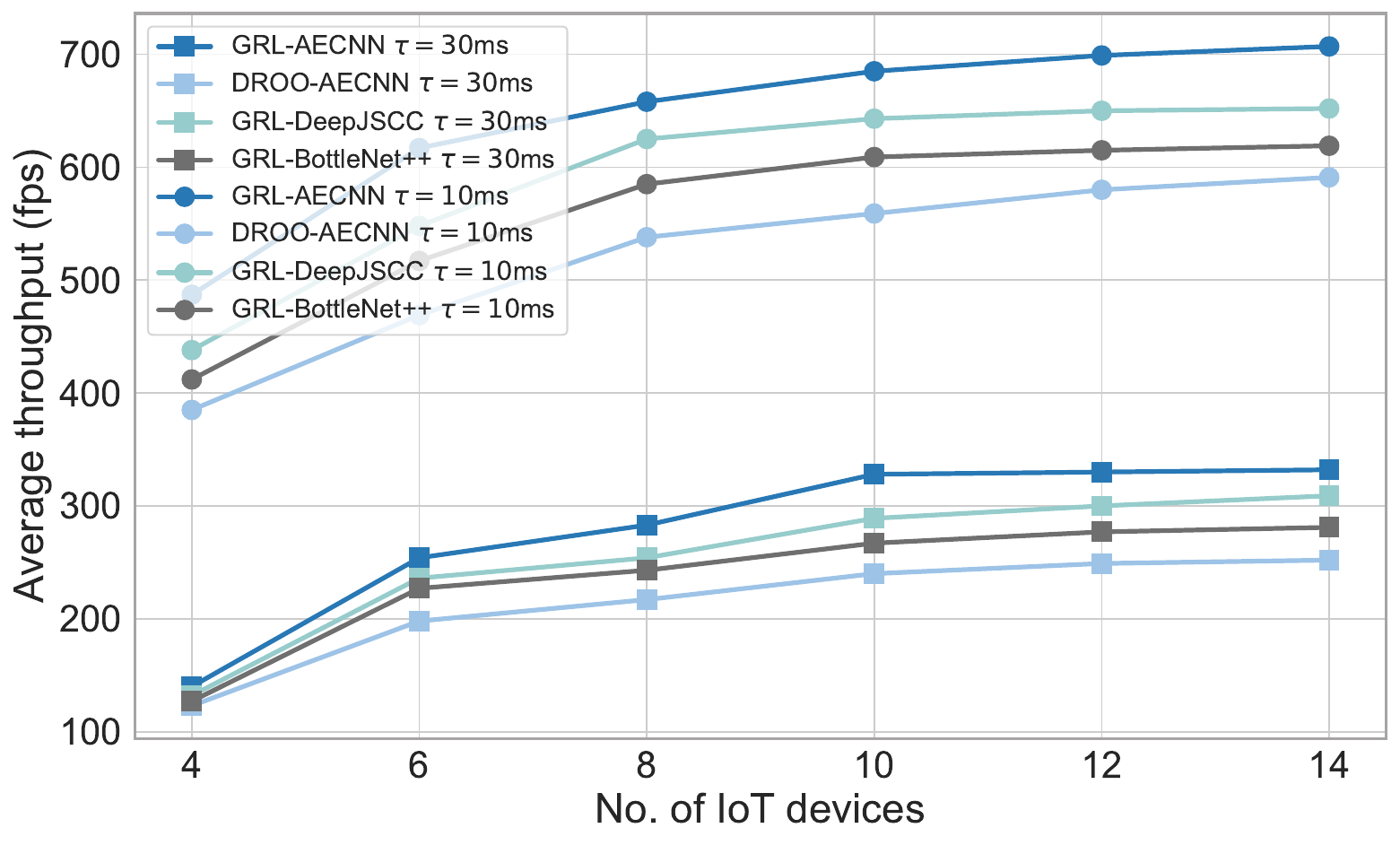}}
    \caption{Performance under various No. of IoT devices}
    \label{fig:iot}
\end{figure*}

\subsection{Performance under Various No. of IoT devices}
We measured the performance of different offloading methods for 10,000 timeslots in different scenarios. We define a successful task as the one that is completed within the deadline and use several metrics to evaluate the reliability, accuracy and efficiency of GRL-AECNN: 
\begin{itemize}
    \item service success probability (SSP): the number of successful tasks divided by the total number of tasks; 
    \item  average inference accuracy: the sum of each successful task's accuracy divided by the total number of tasks; 
    \item average throughput:
    the number of successful tasks is divided by the cumulative number of timeslots. 
\end{itemize}

As shown in Fig. \ref{fig:iot}, the average accuracy and SSP decrease as the number of IoT devices $U$ increases at the given ES computation resources. This is because when $U$ is large, more tasks fail to meet their deadlines due to the limited resources of ESs. As such, the average throughput gradually reaches a plateau. Additionally, we can see that the system achieves a higher throughput at $\tau=10$ ms than that of $\tau=30$ ms; however, the increase in throughput comes with an associated trade-off, i.e., a decrease in both the average accuracy and the SSP. This is because the higher task generation rate at $\tau=10$ ms allows the system to make better use of ES's idle time to process more tasks during the same time duration; however, this will result in more failed tasks because of the higher occupancy of ES and wireless channels. Furthermore, GRL-AECNN demonstrates the capability to enhance average accuracy, SSP, and throughput, particularly in scenarios where $U$ is large. For example, when $U=10$ and $\tau=10$ ms, GRL-AECNN achieves average inference accuracy improvement by 0.20, 0.15, and 0.11 respectively, in comparison with
DROO-AECNN, GRL-BottleNet++, and GRL-DeepJSCC. This is because AECNN can effectively identify the semantic information while reducing the computation on IoT devices via channel pruning, and GRL can use the full information of MEC to make optimal offloading decisions; however, the semantic encoders in BottleNet++ and DeepJSCC lack effective extraction and compression of semantic information while introducing extra computation on IoT devices. 
\begin{figure*}[t]
    \centering
    \subfigure[Average accuracy]{\includegraphics[width=0.45\textwidth]{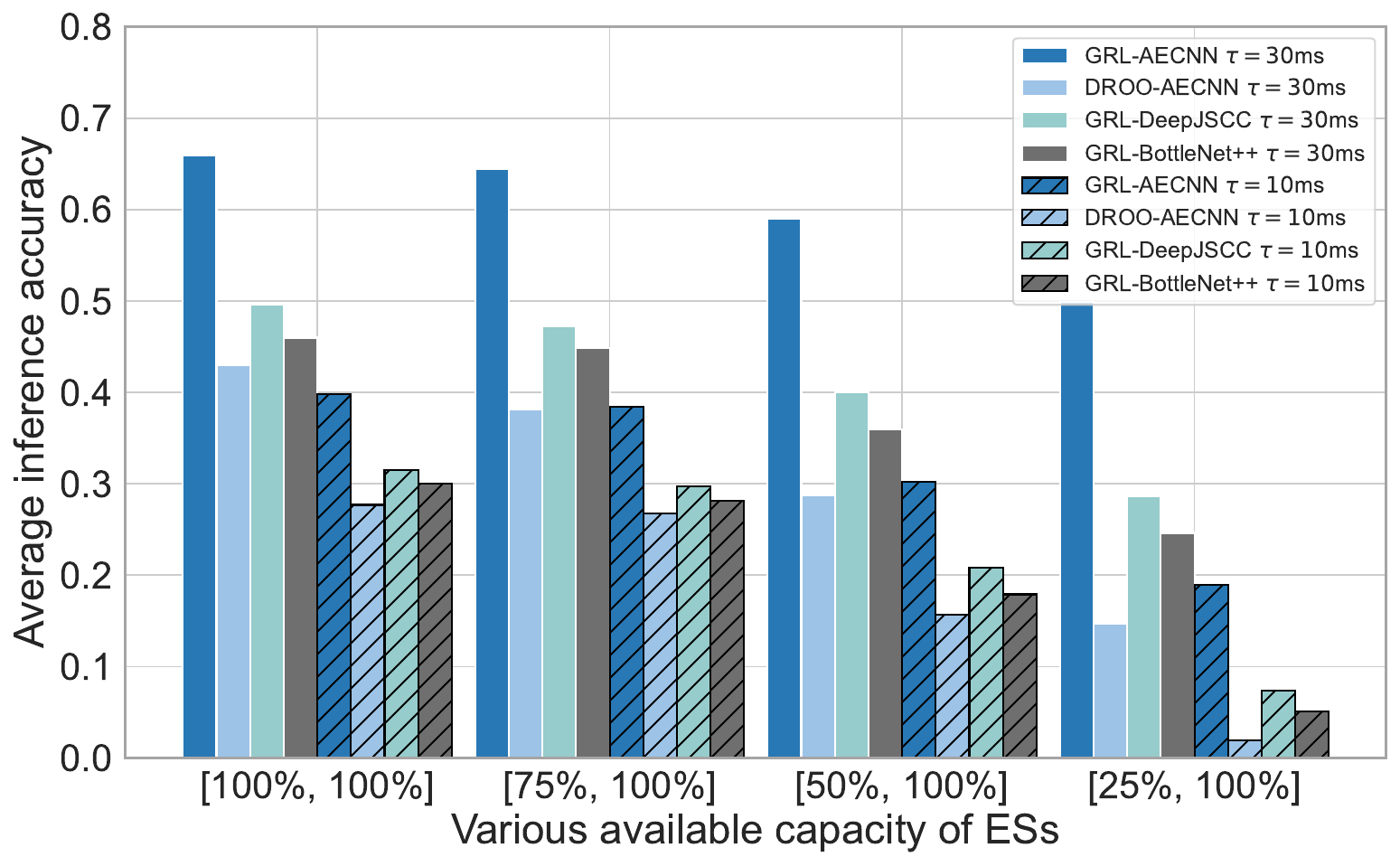}}
    \hspace{5mm}
     \subfigure[Average throughput]{\includegraphics[width=0.45\textwidth]{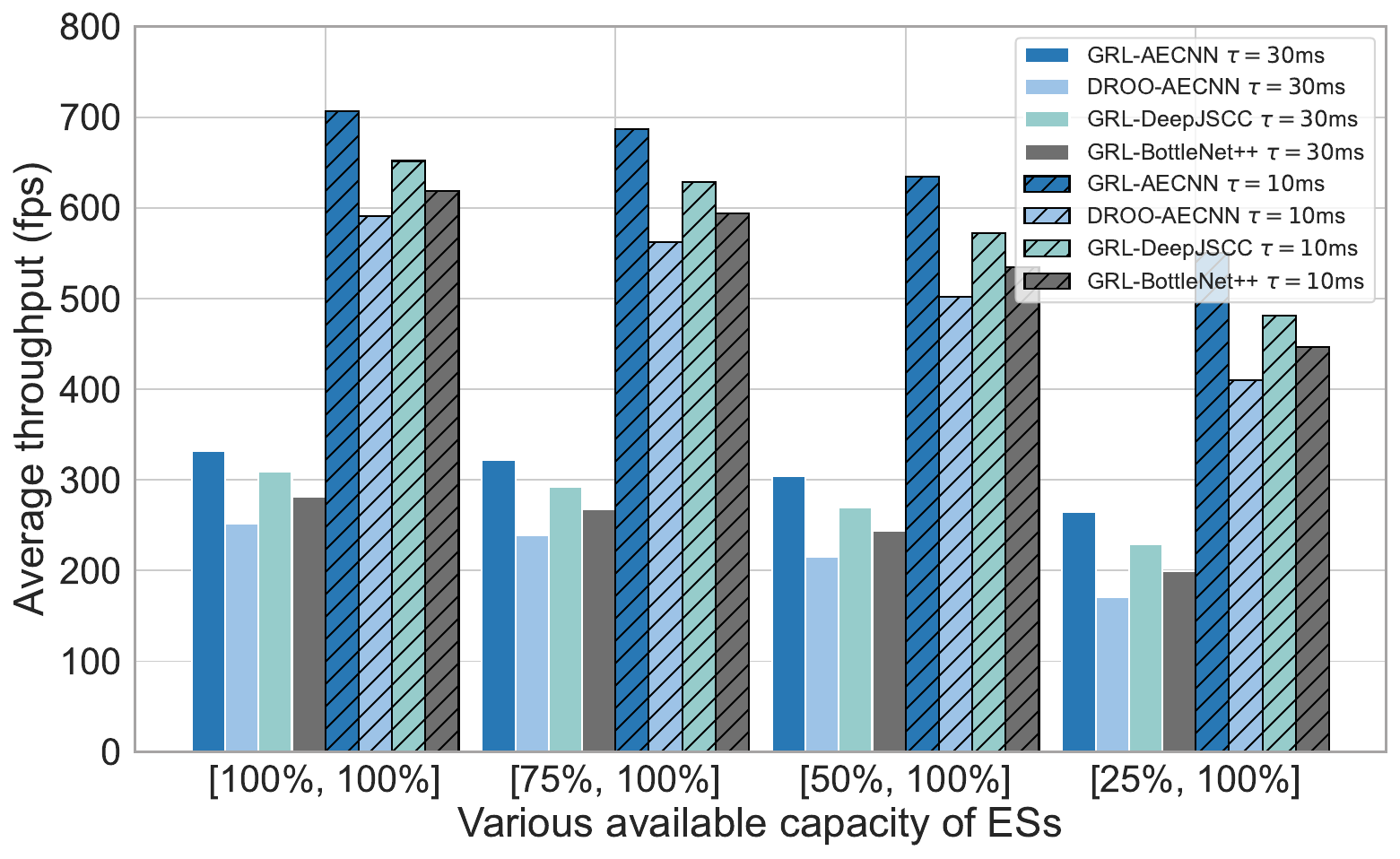}}
    \caption{Performance under various available capacities of ESs}
    \label{fig:ES}   
\end{figure*}

\begin{figure*}[t]
    \centering
    \begin{minipage}[t]{0.45\textwidth}
        \includegraphics[width=\textwidth]{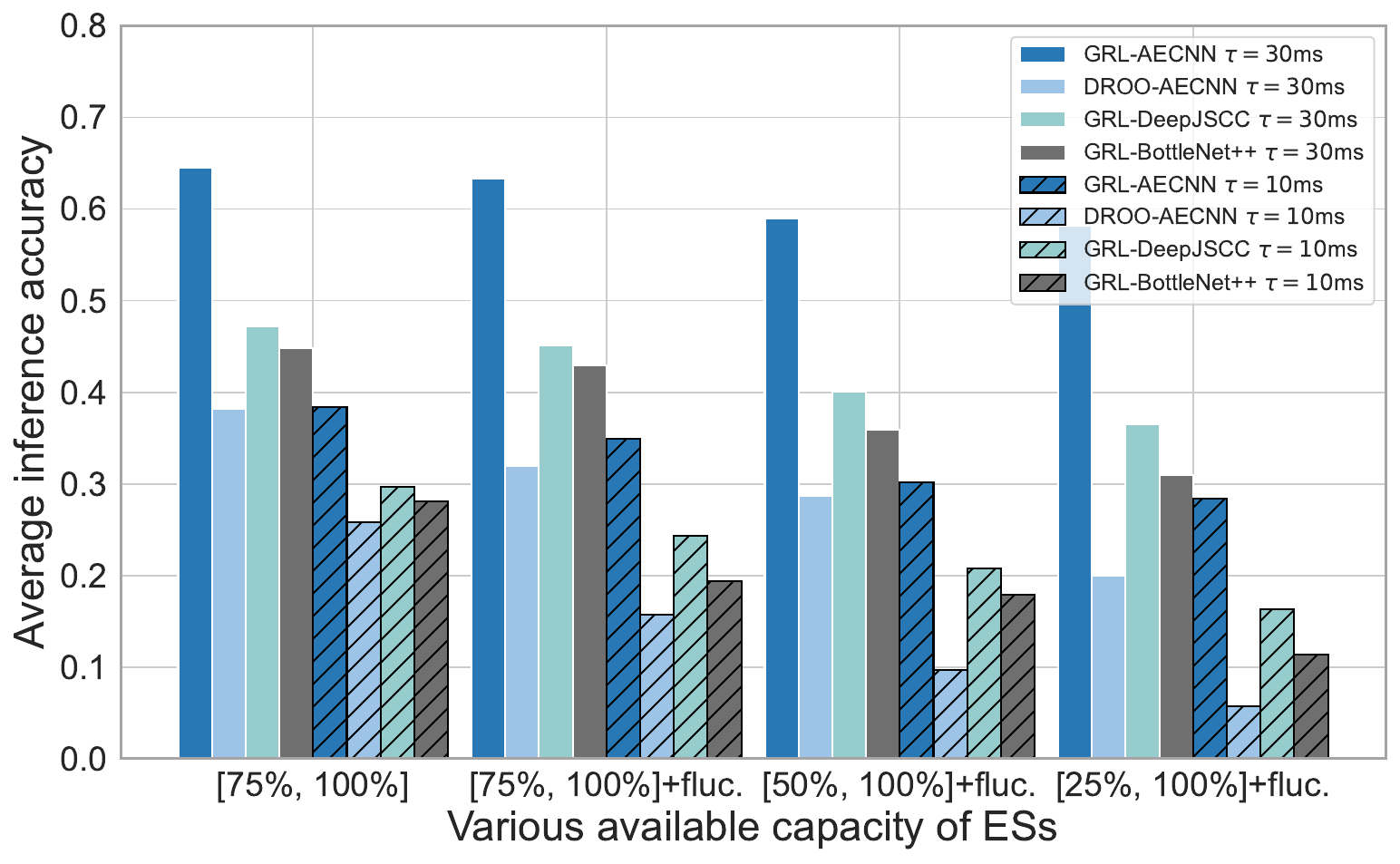}
        \caption{Performance under uncertain computation time}
        \label{fig:fluctuation}
    \end{minipage}
    \hspace{5mm}
    \begin{minipage}[t]{0.45 \textwidth}
    \centering
    \includegraphics[width=\textwidth]{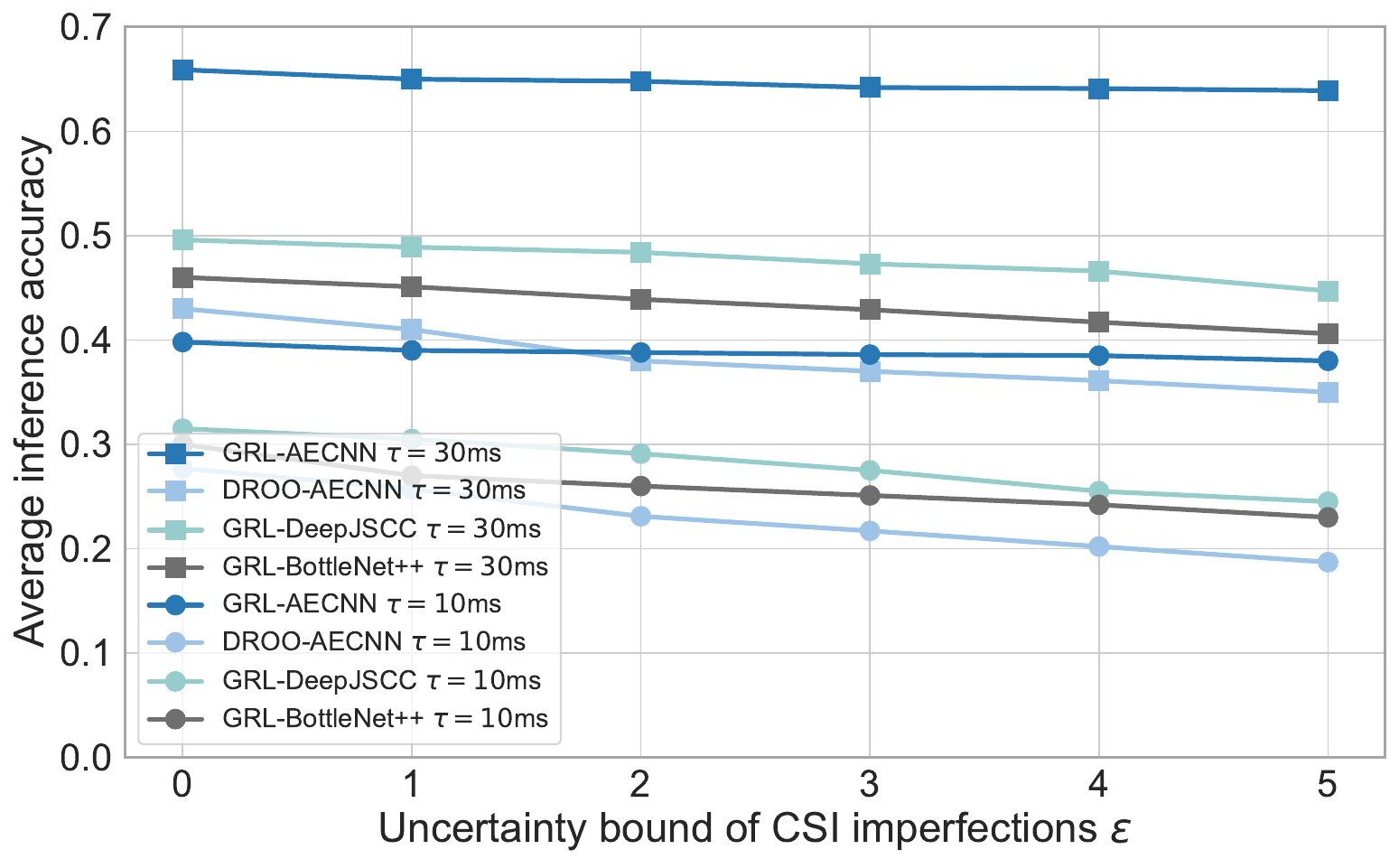}
    \caption{Performance under imperfect CSI}
    \label{fig:csi} 
    \end{minipage}
\end{figure*}
\subsection{Performance under Uncertain Computation Time}
In real-world scenarios, ESs are often not consistently idle and their computational resources are dynamic. To comprehensively reflect the impact of such variations on the effectiveness of offloading strategies, we consider a scenario with 14 IoT devices and 2 ESs where each ES has a stochastic computational resource availability ranging between $\lambda$\% and 100\% of its overall computational capacity at each timeslot. In Fig. \ref{fig:ES}, we compare the performance of the mentioned methods under the case of $\lambda \in \left \{ 25, 50, 75, 100 \right \}$. As the variation range increases, both average inference accuracy and average throughput decrease. This is because insufficient computational resources in ESs make it more likely for tasks to miss deadlines, leading to more task failures, lower average accuracy, and reduced throughput. Notably, GRL-AECNN achieves higher gain over the other methods in terms of average accuracy and average throughput under larger variations in ESs' available computational resources. For example, at $\lambda=25$ and $\tau = 10$ ms, GRL-AECNN improves average inference accuracy up to $(0.189-0.019)/0.019 = 8.9 \times$ over DROO-AECNN and $(0.189 -0.074)/0.074 = 1.6 \times $ over GRL-DeepJSCC respectively, which is higher than that of $(0.302-0.157)/0.157 = 0.9 \times$ and $(0.302-0.208)/0.208 = 0.5 \times$ when $\lambda=75$ and $\tau = 10$ ms. Furthermore, as the variation range increases (i.e., smaller $\lambda$), the three GRL-based offloading methods show lower degradation in average inference accuracy and average throughput than DROO-AECNN. For example, GRL-AECNN has $(0.398-0.189)/0.398 = 0.5 \times$ degradation in average inference accuracy from $\lambda = 100$ to $\lambda = 25$ when $\tau = 10$ ms; however, that of DROO-AECNN up to $(0.277-0.019)/0.277 = 0.9 \times$. This highlights the effectiveness of GRL-AECNN in offloading decision-making for scenarios with limited available computation resources. 

The computation time of ESs can be affected by various factors, including storage availability, thermal conditions, and environmental factors. In addition to the previously mentioned variation in ES computation resources, we considered a realistic case where the computation time of each ES fluctuates by $\pm 25\%$ of its measured value. In Fig. \ref{fig:fluctuation}, the three GRL-based offloading methods remain relatively more stable in average inference accuracy than DROO-AECNN under computation time fluctuations. For example, GRL-AECNN has $(0.384-0.349)/0.349 = 0.1 \times$ degradation in average inference accuracy when $\lambda = 75$ and $\tau = 10$ ms; however, that of DROO-AECNN is $(0.258-0.157)/0.157 = 0.64 \times$. This further demonstrates that GRL-AECNN is better at learning the state information of ESs for effective decision-making.

\subsection{Performance under Imperfect Channel State Information}
Since channel estimation is often not perfect in practical systems, the channel state information (CSI) imperfections can be
deterministically modeled by using the ellipsoidal approximation \cite{Papadaki2008twc}, as
\begin{equation}
    \hat{g}_{u,s}^k = g_{u,s}^k 10 ^{\vartheta_{u,s}^k/10 }, \vartheta_{u,s}^k \in [- \varepsilon, \varepsilon ]. 
\end{equation}
where the non-negative constant $\varepsilon$ denotes the uncertainty bound of CSI imperfections.

In this study, we consider the scenario with 14 IoT devices and 2 ESs, and include the CSI imperfections under different uncertainty bound $\varepsilon$. As shown in Fig. \ref{fig:csi}, the average inference accuracy of the system decreases as the uncertainty bound of CSI imperfections increases. This occurs because the task offloading decisions made at larger biases in imperfect channel estimation might lead to tasks not being completed within their specified deadlines. Consequently, this situation causes more task failures, contributing to an overall decrease in the average inference accuracy. Nonetheless, our proposed GRL-AECNN has a relatively small degradation in terms of average inference accuracy. For example, when $\tau = 10$ ms, GRL-AECNN has $(0.398-0.380)/0.380 = 0.05 \times$ degradation in average inference accuracy from $\varepsilon = 0$ to $\varepsilon = 5$; however, that of DROO-AECNN up to $(0.277-0.187)/0.187 = 0.48 \times$. This highlights the effectiveness of GRL-AECNN in aggregating the CSI imperfections thereby making robust offloading decisions to ensure that more tasks are processed within the deadline in dynamic MEC scenarios.

\section{Conclusion}\label{Conclusion}
In this paper, we studied the computation offloading of CNN inference tasks in dynamic MEC networks. We proposed a novel semantic compression method, AECNN, to address the uncertainties in communication time and available computation resources at ESs. In AECNN, we designed a CA module to figure out the importance of channels, then compressed the intermediate tensor by pruning the less important channels. We used entropy encoding to further reduce communication time by removing redundant information and designed a lightweight CNN-based FR module to recover the intermediate tensor through learning from the received compressed tensor to improve accuracy. We designed a reward function to trade off the inference accuracy and task completion time, and formulated the CNN inference offloading problem as a maximization problem to maximize the average inference accuracy and throughput in the long term. To address the optimization problem, we proposed GRL-AECNN to make the optimal offloading decision and use a step-by-step approach to fasten the training process. The experimental results show that GRL-AECNN can achieve better performance in terms of average inference accuracy, service success reliability, and average throughput than the state-of-the-art methods, which highlights the effectiveness of GRL-AECNN in aggregating all
the information of dynamic MEC thereby making robust offloading decisions.
\ifCLASSOPTIONcaptionsoff
  \newpage
\fi

\bibliographystyle{IEEEtran}
\bibliography{main}

\appendices

\renewcommand\thefigure{\Alph{section}\arabic{figure}}  
\renewcommand\theequation{\Alph{section}\arabic{equation}}  
\setcounter{equation}{0}
\section{FLOPs Count in CNN} \label{Flops}
In CNNs, the input feature map of a CL is derived from the output feature map of its preceding CL. For instance, let's consider CL $l$, which takes the input feature map denoted as $\mathcal{X}_{l-1} \in \mathbb{R}^{C_{l-1} \times H_{l-1} \times W_{l-1}}$ and produces the output feature map $\mathcal{X}_{l} \in \mathbb{R}^{C_{l} \times H_{l} \times W_{l}}$. The computation of FLOPs for CL $l$ captures the computational complexity associated with processing data through CL $l$. 

According to the work \cite{035}, the FLOPs calculation of CL $l$ is primarily influenced by the dimensions of the input and output channels, the kernel size employed by the convolution operation, and the dimensions of the resulting output height and width. As such, the FLOPs for CL $l$ are calculated using the following formula from \cite{035}:
\begin{equation}
    \begin{array}{*{20}{c}}
    \xi_l = C_{l-1} C_l f_l^2 H_l W_l,& \forall l \in \mathcal{L},
    \end{array}
\end{equation}
where $f_l$ is the kernel size of CL $l$.

Regarding the FC layer, the computational operations primarily consist of multiplying the input data by weight parameters and then applying activation functions. We consider an FC layer $l'$ with the input feature map $\mathcal{X}_{l'}^{\textit{in}} \in \mathbb{R}^{1 \times D_{l'}^{\textit{in}}}$ and output feature map $\mathcal{X}_{l'}^{\textit{out}} \in \mathbb{R}^{1 \times D_{l'}^{\textit{out}}}$. According to \cite{036}, the FLOPs of an FC layer $l'$ on can be calculated as 
\begin{equation}
	\xi_{l'} = D_{l'}^{\textit{in}} D_{l'}^{\textit{out}},
\end{equation}
where $D_{l'}^{\textit{in}}$ and $D_{l'}^{\textit{out}}$ are the input dimensionality and the output dimensionality of FL ${l'}$.

\section{Proof of Theorem \ref*{theorem1}} \label{proof}

To prove Theorem \ref*{theorem1}, we first introduce the hyperbolic tangent function ${\textit{sigmoid}}(x) = \frac{e^x }{e^x + 1}$. The function ${\textit{sigmoid}}(x) $ is a nonlinear function with values between 0 and 1 \cite{Giovanni2017}. As shown in Fig. \ref{fig:sigmoid}, ${\textit{sigmoid}}(x)$ approaches $1$ as $x \to 5$, while ${\textit{sigmoid}}(x)$ approaches $\frac{1}{2}$ as $x \to 0$. 

\setcounter{figure}{0}  
\begin{figure}[h]
    \centering
    \includegraphics[width=0.45\textwidth]{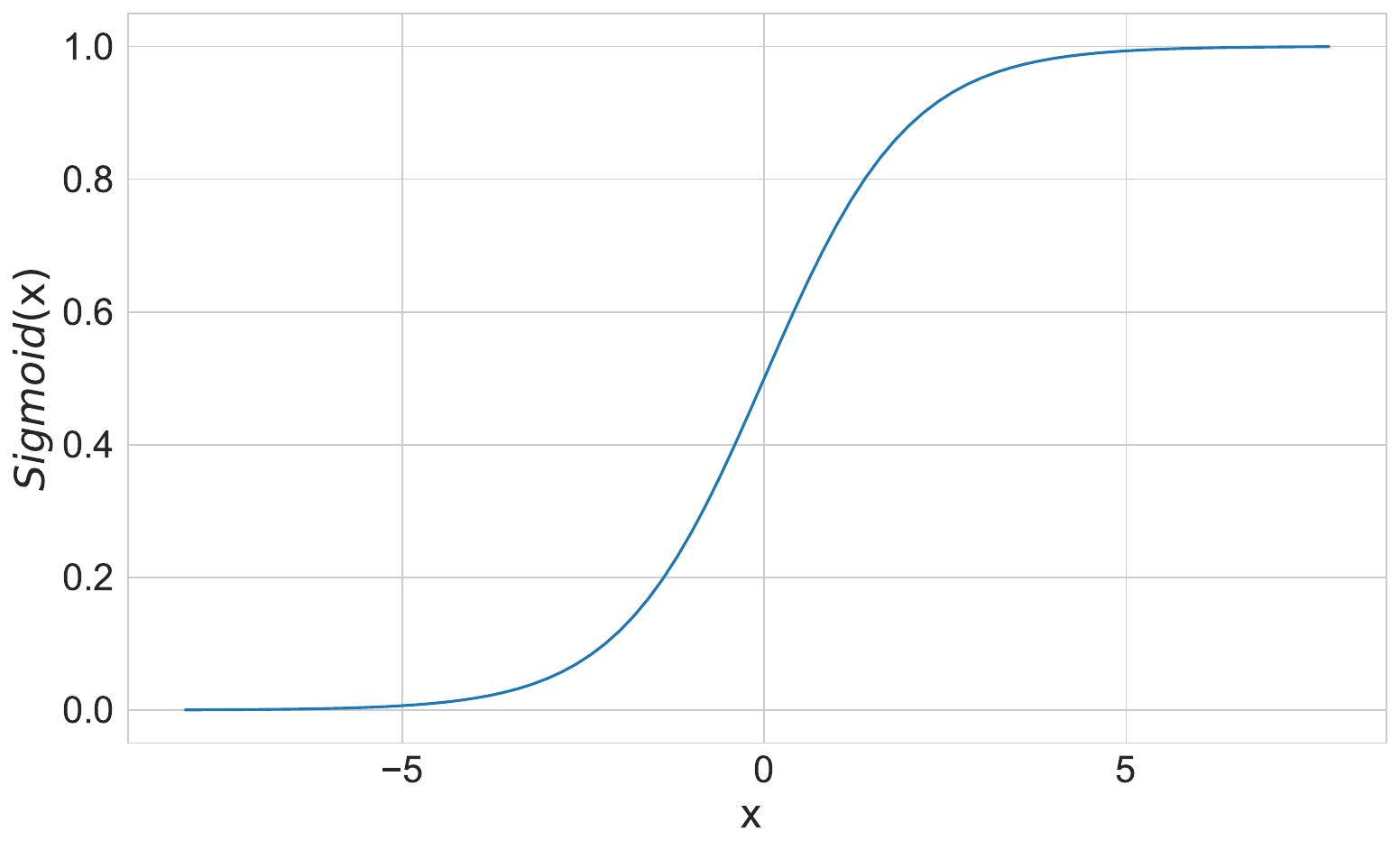}
    \caption{$\textit{sigmoid}(x)$ function}
    \label{fig:sigmoid}
\end{figure}
Based on the above, we define a new function $\psi'(x) = 2\left(1-{\textit{sigmoid}}(x)\right)$ and derive that $\psi'(x)$ approaches 0 as $x$ approaches 5 and approaches 1 as $x$ approaches 0. Extending this to our original function $\psi(x) = 2\left(1 - {\textit{sigmoid}}\left(\frac{5x}{\sigma_u^k}\right)\right)$, we can deduce that $\psi(x)$ behaves similarly: as $x$ approaches $\sigma_u^k$, $\psi(x)$ approaches 0; and as $x$ approaches 0, $\psi(x)$ approaches 1.

\vfill

\end{document}